\catcode`\@=12

\documentstyle[12pt]{article}
\newcommand{\bea}{\begin{eqnarray*}}
\newcommand{\eea}{\end{eqnarray*}}
\newcommand{\BE}{\begin{eqnarray}}
\newcommand{\EN}{\end{eqnarray}}
\newcommand{\be}{\begin{equation}}
\newcommand{\en}{\end{equation}}

\newcommand{\no}{\noindent}
\newcommand{\vs}{\vspace}

\begin{document}

\title{ Science of Chaos or Chaos in
Science?\thanks{To appear in Physicalia Magazine, and in the
Proceedings of the New York Academy of Science. }}

\author{J.Bricmont \\
Physique Th\'eorique, UCL,\\
B-1348 Louvain-la-Neuve, Belgium
 }

\date{}
\maketitle
\begin{abstract}

 I  try to clarify
several confusions
in the popular literature concerning chaos, determinism, the arrow
of time, entropy and the role of probability in physics.
Classical ideas going back to
Laplace and Boltzmann are
explained and defended
while some recent views on irreversibility, due to
Prigogine, are criticized.

\end{abstract}

\section{Introduction}
\begin{flushleft}
{\footnotesize We might characterize today's breakdown of
industrial or ``Second Wave"\\
 society as a civilizational ``bifurcation", and the rise
of a more differentiated, \\
``Third Wave" society as a leap to new
``dissipative structures" on a world scale.\\
 And, if we accept this analogy, might we not
 look upon the leap from Newtonianism to \\
Prigoginianism in the
same way? Mere analogy, no doubt. But illuminating, nevertheless.\\
Alvin Toffler (preface to \cite{PS3}).} \end{flushleft}

\vs{3mm}

Popularization of  science seems to be doing very well:
the Big Bang, the theory of elementary particles or of black holes
are explained in countless books for the general public.
The same is true for chaos theory, irreversibility or self-organization.
However, it seems also that a lot of confusion exists concerning these
latter notions, and that at least some of the popular books are
 spreading misconceptions. The goal of this article is to examine
some of them, and to try to clarify the situation.

In
particular, I will make a
 critical evaluation of the various claims
concerning chaos and irreversibility made by Prigogine
and by Stengers, since ``La Nouvelle
Alliance". Several of those claims, especially the most recent
ones, are rather radical:
 ``the notion of chaos leads us to rethink the notion of ``law of
nature"." (\cite{P2}, p.15)\footnote{Here and below, I have translated
the texts that were available only in French.} For chaotic systems,
``{\it trajectories are eliminated from the probabilistic
description} \dots The statistical description
 is {\it irreducible}." (\cite{P2}, p.59) The
existence of chaotic dynamical systems supposedly
 marks a radical departure from a fundamentally
deterministic world-view, makes  the
notion of trajectory obsolete, and offers a new
understanding of irreversibility.
 Prigogine
and Stengers claim that the classical
 conception was unable to incorporate time in our
view of the world (\cite{PS2}, chap.1)
or to account for the irreversibility of macroscopic
phenomena. Boltzmann's attempt to explain irreversibility on the basis
of reversible laws failed (\cite{P2}, p.41).

On the basis of these theories, a  number of
speculations are put forward on the notion of ``event", on
the place of human beings in Nature,
 or
even on overcoming Cartesian dualism (see
\cite{P2}, chap.9, \cite{P3}, p.106, and \cite{P4}).
These writings have been indeed quite influential, mostly among
non-experts. They are frequently quoted in philosophical or cultural
circles, as an indication that chaos, nonlinear phenomena or the ``arrow
of time" have led to a profound revolution in our way of thinking.

I want to develop quite different views on most of these issues. In my opinion,
chaos does not invalidate in the least the
classical deterministic world-view;
the existence of
chaotic dynamical systems actually strengthens that view (Sect. 2).
Besides, the relationship between chaos
and irreversibility is quite different from what
is claimed e.g. in ``Les lois du chaos" \cite{P2} . Finally,
when they are correctly
presented, the classical views of  Boltzmann perfectly account for
macroscopic irreversibility on the basis of deterministic, reversible,
microscopic laws (Sect. 3). Part of the difficulty in understanding those views comes
from some confusions about the use of the words ``objective" and
``subjective", associated with probability or entropy. I will try
to be careful with these notions (Sect. 4 and 5).
In section 6, I will discuss the applications of probabilistic reasoning to
complex phenomena and biology.
I shall also argue that most of the speculation on
the ``new alliance" between the human sciences and the natural ones is misguided and
that the people working in sociology or psychology have very
little to learn from the alleged ``leap from Newtonianism to
Prigoginianism"
 (Sect. 7).

On the other hand, I believe that the  ideas
of Laplace and of Boltzmann  are worth defending against various
misrepresentations and misunderstandings. Quite independently of the work
of Prigogine, there are serious confusions that are found in the
literature on irreversibility,
chaos or time (some of which go back to philosophers such  as
Popper, Feyerabend or Bergson). Besides, many textbooks or popular
books on statistical
mechanics are rather
obscure, at least in the part concerning the
foundations of the field (e.g., on the role played by ergodic theorems).
 I will try to clarify these questions too (Sect. 4).

I  wrote this paper  in a not too technical language,
relegating formulas to footnotes
and remarks. Nothing of what I say is new\footnote{On the issue
of irreversibility, see Feynman \cite{Fe2},
Jaynes \cite{Ja}, Lebowitz \cite{Le1,Le2}, Penrose \cite{Pe2}. For a similar
and less technical
critique of various confusions about chaos, see Maes \cite{Mae}.}. In
fact, everything is quite standard and old, and it is a sad fact that those
ideas that were so nicely explained by Boltzmann  a century ago
\cite{Bo} have to be reexplained over and over again.

Finally, I have to emphasize that this is in no way
a criticism of Prigogine's work in general,
and even less of the Brussels' school. I shall only discuss
the radical claims made in the popular books and, in particular, the idea
that   fundamental flaws have been found
in the scientific world-view
and that one has to rethink the {\it notion} of law of nature.
 I believe that a
lot of interesting scientific ideas have been
developed around Prigogine and that he has had an
exceptional taste for discovering new directions
 in physics, whether in irreversible
thermodynamics or in chaotic phenomena. But this does
not put his views on foundational questions
beyond criticism\footnote{I must add
that I have defended, in the past, some of the ideas criticized below.
Needless to say, I am interested in the critique of ideas and not of
individuals.}.

\section{Chaos and determinism:  Defending  Laplace.}
\begin{flushleft}
{\footnotesize The concept of dog does not bark.\\
B. Spinoza} \end{flushleft}

\vs{3mm}

\no {\bf 2.1. Determinism and predictability.}

\vs{3mm}

A major scientific development in recent decades has been popularized
under the name of ``chaos". It is widely believed that this
implies a fundamental philosophical or conceptual revolution.
In particular, it is thought that the
classical world-view brilliantly expressed
by Laplace in his ``Philosophical Essay
on Probabilities" has to be
 rejected\footnote{ For rather negative comments
on Laplace,
see e.g. Ekeland (\cite{Ek}, p.31), and
Gleick (\cite{Gl}, p.21 in the French edition).}.
Determinism is no longer
defensible.
I think this is based on a serious confusion between {\it determinism}
and {\it predictability}. I will start by underlining the difference
between the two concepts. Then, it will be clear that what goes under the name
 of ``chaos" is a major scientific progress but does not have the radical
philosophical implications that are sometimes attributed to it.

In a nutshell,
determinism has to do with how Nature behaves,
and predictability is related to what we,
human beings, are able to observe, analyse and compute.
It is easy to illustrate the necessity for such a
distinction. Suppose we consider a perfectly
regular, deterministic {\it and} predictable mechanism,
like a clock, but put it on the top of a
mountain,
or in a locked drawer, so that its state (its initial
conditions) become inaccessible to us.
This renders the system trivially unpredictable, yet it seems
difficult to claim that it becomes
non-deterministic\footnote{Likewise, only the most radical
social constructivist might object to the idea
that Neptune or Pluto were following their
(deterministic) trajectories before they were
discovered.}. Or consider a pendulum: when there is no external force, it
is deterministic and predictable. If one applies to it a periodic forcing, it
may become unpredictable. Does it cease to be deterministic?

In other words, anybody who admits that {\it some} physical phenomena
obey deterministic laws must also admit that some physical phenomena, although
deterministic, are not predictable, possibly for ``accidental" reasons.
So, a distinction must be made\footnote{In an often quoted lecture to the
Royal Society, on the three hundredth anniversary of Newton's Principia, Sir James Lighthill
gave an inadvertently perfect example of how to slip from  inpredictability
to indeterminism: "We are all deeply
conscious today that the enthusiasm of our forebears for the
marvellous achievements of Newtonian
mechanics led them to make generalizations in this area of
{\it predictability} which, indeed, we may
have generally tended to believe before 1960, but which we now recognize were false. We
collectively wish to apologize for having misled the general
educated public by spreading ideas
about {\it determinism} of systems satisfying Newton's laws of
motion that, after 1960, were to be
 proved incorrect..." \cite{Li} (Italics are mine;
quoted e.g. by Reichl \cite{Re}, p.3,
and by Prigogine and Stengers, \cite{PS2}, p.93, and \cite{P2}, p.41).
See also, \cite{Va} p.7 where, after describing a chaotic system,
one concludes that ``the deterministic approach fails"
.}. But, once this is admitted, how does one show
that {\it any} unpredictable system
is {\it truly}
non-deterministic, and that the lack of predictability is not merely due
to some limitation of our abilities? We can never infer indeterminism from
our ignorance alone.

Now, what does one mean exactly by determinism?
Maybe the best way to explain it is to go back to Laplace
:`` Given for one instant an intelligence which could comprehend
all the  forces by which nature is animated and
the respective situation of the beings who compose
it- an  intelligence sufficiently vast to submit
these data to analysis- it would embrace in the
same formula the movements of the greatest bodies
of the universe and those of the lightest atom;
for it, nothing would be uncertain and the future,
as the past, would be present before its eyes."
\cite{La} The idea expressed by Laplace
 is that determinism depends on what the laws
of nature are. Given the state of the system at some time,
we have  a formula
(a differential equation, or a map) that
gives in principle
the state of the system at a later time.
To obtain predictability, one has to be able
to measure the present
state of the system with enough precision, and
 to compute with the given formula
(to solve the equations of motion). Note that there exist alternatives
to determinism: there could be no law at all; or the laws could be stochastic:
the state at a given time (even if it is known in
every conceivable detail) would determine only a probability distribution
for the state at a later time.

How do we know whether determinism is true, i.e. whether nature
obeys deterministic laws?
This is a very complicated issue.
Any serious discussion of it must be based on an
analysis of the fundamental laws, hence of quantum
mechanics, and I do not want to enter this
debate here\footnote{
I have expressed my point of view
on the foundations of quantum mechanics in
\cite{Br}; for related views, see Albert \cite{Al,Al1},
Bell \cite{Be}, D\"urr et al. \cite{DGZ},
Maudlin \cite{Ma}. For a precise discussion of
determinism in various physical theories, see Earman \cite{Ea}.
\protect\label{note_QM}}.
Let me just say that it is conceivable that we shall
obtain, some day,
a complete set of fundamental physical laws (like the law
of universal gravitation
in the time of Laplace), and then, we
shall see whether these laws are deterministic or
not\footnote{Most of the laws that are discussed in the
literature on chaos (e.g. on the weather)
are actually macroscopic laws, and
not fundamental, or microscopic ones.
This distinction will be discussed in Section 3.}. Any discussion of
determinism outside of the framework of the fundamental laws
is useless\footnote{Opponents of determinism
are quick to point out that determinism cannot be proven. While it is
of course true that no
statement about the world can literally be {\it proven}, these
opponents do not always
see how vacuous are their own arguments in favour of
 indeterminism, arguments that rely ultimately
on our ignorance. Popper, in \cite{Pop5}, gives
 a long series of such arguments. In a review of this book, the biologist
Maynard Smith shows a rather typical misunderstanding of Laplace:
first, he agrees with Popper's criticism of Laplace, because the
latter's computations are, in practice, impossible to do.
But, then he  disagrees with Popper
about free will
and gives, as far as I can see,
 a perfectly causal and Laplacian account of human
actions, which of course, are not practically
 computable either (\cite{May}, p.244).
To avoid misunderstandings, I am not trying to say that determinism
is or must be true. All I say is that various arguments against determinism
miss the point.}.
 All I want to stress here is that
the existence of chaotic dynamical systems does not
affect {\it in any way} this discussion. What are chaotic systems?
The simplest way to define them is through
sensitivity to initial conditions. This means that,
for any initial condition
of the system, there is some other initial condition,
 arbitrarily close to the first
one so that, if we wait long enough, the two systems will be markedly
different\footnote{Here is a simple example. Consider the ``phase space"
to be simply the interval $I=
[0,1[$. And take as (discrete time) dynamics
 the map $f: x \rightarrow 10x$ $mod$ $1$.
This means, we take a number between $0$ and $1$, multiply it by $10$,
write the result as an integer plus a number between $0$ and $1$  and take
the latter as the result (i.e. $f(x)$).
This gives again a number between
$0$ and $1$, and we can repeat the operation. Upon iteration, we obtain
 the {\it orbit} of $x$; $x$ itself is the initial
condition. To describe concretely the latter,
one uses the decimal expansion. Any number in $I$ can be written as
$x=0.a_1a_2a_3\dots$, where $a_i$ equals $0,1,2,\dots,9$. It is easy to see
that $f(x)=0.a_2a_3\dots$. This is a perfect example of a {\it deterministic}
but {\it unpredictable} system. Given the state $x$ at
some initial time, one has a rule
giving the state of the system for arbitrary times. Moreover, for any
fixed time, one can, in principle, find the state after that time, with
any desired accuracy, given a sufficiently precise characterization
of the initial state. This expresses the deterministic aspect.
Unpredictability comes from the fact that, if we take two initial
conditions at a distance less than
$10^{-n}$, then the corresponding orbits could
differ by, say, $1/2$, after n steps, because the difference will be
determined by the $n$th decimal. One of the relatively
recent discoveries in dynamical systems
is that simple physical examples, like a forced pendulum, may behave
more or less like this map.\protect\label{note_10x}}. In other words,
an arbitrarily small error on
the initial conditions makes itself felt after a long enough time.
Chaotic dynamical systems are of course unpredictable in practice, at least
for long enough times\footnote{How long a time this is  depends
 on the details of the system.}, since there will always be some error
in our measurement of the initial conditions. But this does not have any
impact on our discussion of determinism, since we are assuming from
the beginning that the system obeys some deterministic law. It is only
by analysing this
deterministic system that one shows that a
small error in the initial conditions may lead
to a large error after some time.
If the system did not obey any law, or if it followed a stochastic law, then
the situation would be very different. For a stochastic law,
two systems with the {\it same} initial
condition could be in two very different states after a short
time\footnote{It is worth observing that Turing machines, or the Game
of Life, provide examples of deterministic automata whose evolution
is more unpredictable (in a precise technical sense)
 than the one of the usual chaotic dynamical systems.}.

It is interesting to note that the notion that
 small causes can have big effects (in a perfectly
deterministic universe) is not new at all. Maxwell
wrote: ``There is a maxim which
is often quoted, that ``The same causes will always produce the same
effects"". After discussing the meaning of this principle, he adds: ``There is another maxim
which must not be confounded with that quoted at the beginning of this article, which asserts
``That like cause produce like effects."
This is only true when small variations in the initial circumstances produce only small
variations in the final state of the system"(\cite{Max1}, p.13)\footnote{As
for applications to the weather, Poincar\'e (\cite{Po}, p.69)
already noticed that the
 rainfalls or the storms seem to occur at random,
so that people are more likely to pray
for rain than for an eclipse (for an exception to this rule, but
based on prior knowledge, see \cite{Ti}, p.59). We are
not able to predict the storms, because the atmosphere may be
 in a state of ``unstable equilibrium".
It may all depend on a tenth of a degree. And he adds: ``If we had known this tenth of a degree,
one  could have made predictions", but since our observations are not sufficiently precise, it
all appears to be due to randomness.
}.
One should not conclude from these
quotations\footnote{Hadamard, \cite{Ha}, Duhem \cite{Du} and
Borel \cite{Bor} made similar observations. See Ruelle, \cite{Ru}
for a discussion of that history from a modern perspective, and
a good popular exposition of chaos. }
that there is nothing new under the sun. A lot more is
known about dynamical systems than in the time of Poincar\'e. But, the
general idea that not everything is predictable,
 even in a deterministic universe, has
been known for centuries.
Even Laplace emphasized this point: after
formulating universal determinism, he stresses
that we shall always remain
``infinitely distant" from the intelligence that he just
introduced. After all, why is this determinism stated in a
 book on {\it probabilities}?
The reason is obvious: for Laplace,
probabilities lead to rational inferences in situations
of incomplete knowledge (I'll come back below
to this view of probabilities).
So he is assuming from the beginning that our knowledge is incomplete, and
that we shall never be able to {\it predict} everything. It is a complete mistake
to attribute to some ``Laplacian dream" the idea of
perfect predictability\footnote{ It is interesting to
read the rest of the text of Laplace.
 First of all,
he   expresses the belief that there are indeed fundamental, universally
valid laws of nature which can be discovered through scientific
investigation (the only examplethat Laplace had of a fundamental law
was that of universal gravitation). In that respect,
nothing has changed today. One of the goals of  physics is still to
discover those fundamental laws. His basic idea could be called universal
reductionism rather than universal determinism.
Since reductionism is remarkably well defended in Weinberg's book
``Dreams of a Final Theory" \cite{We}, I shall not pursue this point.
Reading a little further, we see that Laplace's goal is to use
science against superstition. He mentions the fears caused by Halley's comet
in the Middle Ages (where it was taken as a sign of the divine wrath)
and how our discovery of the laws of the ``system of the world" ``dissipated
 those childish fears due to our ignorance of the true relations between Man
and the Universe".
Laplace  expresses also a deep optimism about the progress of
science. Again nothing of that has been refuted by the evolution of natural
sciences over the last two centuries.
But one will not find any claim about
the computability, by us humans, of {\it all} the consequences of the laws of
physics.}.
But  Laplace does not commit  what
 E. T. Jaynes calls
the ``Mind Projection Fallacy": ``We are all under an
ego-driven temptation to project our private thoughts
out onto the real world, by supposing
that the creations of one's own imagination are real
properties of Nature, or that one's own
ignorance signifies some kind of indecision on the part
of Nature" \footnote{Jaynes'
criticisms were mostly directed  at the way quantum
theory is presented, but they also apply
to some discussions of chaos theory or  of
statistical mechanics.}(\cite{Ja2}, p.7). As we shall see, this is a most
common error. But, whether we like it or not, the concept of dog does not
bark, and we have to carefully distinguish between
 our representation of the world
and the world itself.

Let us now see why the existence of chaotic dynamical systems in fact
supports universal determinism rather than contradicts it\footnote{Of course,
since classical mechanics is not really  fundamental (quantum mechanics is),
 this issue is rather
academic. We nevertheless want to discuss it, because there seems
to be a lot of confusion in the literature.}. Suppose for a moment that
 no classical mechanical system can behave chaotically. That
is, suppose we have a theorem saying that any such system must eventually
behave in a periodic fashion\footnote{Imagine, for example, that the Poncar\'e-Bendixson
theorem
 held in all dimensions.}. It is not completely obvious what the
conclusion would be, but certainly {\it that} would be an embarassment for
the classical world-view. Indeed, so many physical systems seem to behave in
a non-periodic fashion that one would be tempted to conclude that classical
mechanics cannot adequately describe those systems. One might suggest that
there must be an inherent indeterminism in the basic laws of nature. Of
course, other replies would be possible: for example, the period of those
classical motions might be enormously long. But it is useless to speculate on
this fiction since we know that chaotic behaviour is compatible with a
deterministic dynamics. The only point of this story is to stress that
deterministic chaos increases the explanatory
power of deterministic assumptions, and therefore, according to normal
scientific practice, {\it strengthens} those assumptions.  And, if we did not
know about quantum mechanics, the recent discoveries about chaos would not
force us to change a single word of what Laplace wrote\footnote{And, concerning
quantum mechanics, see the references in note \protect\ref{note_QM}.}.

\vs{3mm}

\no {\bf 2.2. Trajectories and probabilities.}

\vs{3mm}

Now, I will turn to  the main thesis
of Prigogine and his collaborators on chaotic dynamical systems: the notion of
trajectory should be abandoned, and replaced by probabilities.
What does this mean?
Let me quote Prigogine: ``Our leitmotiv is that the formulation
of the dynamics for chaotic systems must be done at the
probabilistic level" (\cite{P2}, p.60). Or: `` We must therefore
eliminate the notion of trajectory
from our microscopic description. This actually corresponds to a
realistic description: no measurement, no computation
lead strictly to a point, to the consideration of a
{\it unique} trajectory.
We shall always face a {\it set} of trajectories"
(\cite{P2}, p.60)\footnote{See also
  \cite{PS2}, p.28: ``As we shall see,
there exists, for sufficiently unstable systems, a
``temporal horizon" beyond which
no determined trajectory can be attributed to them".
To be fair, I should add that
these radical statements are combined with more reasonable, but more
technical ones, e.g.: ``This formulation implies that one must study the
eigenfunctions and the eigenvalues of the
evolution operator." (\cite{P2}, p.60) The question is of course which
statements will
strike most the non-specialized reader.}.

 Let us first see how reasonable
it is to ``eliminate the notion of
trajectory" for chaotic systems by considering
  a concrete example\footnote{
See Batterman
\cite{Ba} for a related, but different, critique. Batterman says that
the replacement of trajectories
by probabilities is ``very much akin to the claim in quantum mechanics
 that the
probabilistic state description given by the $\Psi$-function
is complete, that is,
that underlying exact states cannot exist." (p.259)
But he notes that here, unlike in quantum mechanics,
no no-hidden variable argument is given to
support that claim (for the exact
status of no-hidden variable arguments in quantum mechanics,
see Bell, \cite{Be} and Maudlin, \cite{Ma}).}. Take a billiard ball
on a sufficiently smooth table, so that we can neglect friction (for some
 time), and assume
that there are suitable obstacles and boundaries so that the system is
chaotic. Now suppose that we use an ``irreducible" probabilistic description,
that is, instead of assigning a position to the ball, we assign to it
a probability  distribution\footnote{An absolutely continuous one, i.e.,
given by a density. If one considers probabilities given by delta functions, it
is equivalent to considering trajectories.\protect\label{note_proba}}. Consider
next the evolution of that probability
distribution. Since we are dealing with a chaotic system, that distribution
will spread out all over the billiard table. This means that after a rather
short time, there will be an almost uniform probability of finding the ball
in any given region of the table. Indeed, even if our initial probability
distribution is well peaked around the initial position of the ball, there
will be lots of nearby initial conditions that will give rise to very
different trajectories (that is exactly what it means to say
that the system is
chaotic).  But now we can hardly
take the probability distribution after some time seriously as an
``irreducible" {\it description} of the system. Indeed, whenever we look at
the system, we find the ball somewhere, at a rather well defined position on
the table. It is certainly not completely described by its probability
distribution. The latter describes adequately our knowledge (or rather our
ignorance) of the system, obtained on the basis of our initial information.
But it would be difficult to commit the Mind Projection Fallacy more
radically than to confuse the objective position of the ball and our best
bet for it . In fact, chaotic systems  illustrate this difference:
if all nearby initial conditions followed nearby trajectories, the distinction
between probabilities and trajectories would
not matter too much. But chaotic system show exactly how unreasonable
is the assignment of ``irreducible" probabilities, since the latter quickly
spread out over the  space in which the  system evolves.

Of course, nobody will deny that the ball is always somewhere. But this example
raises the following the question: what does it mean exactly to ``eliminate
trajectories"\footnote{And somewhat more importantly, what does the general educated
public, which reads the popular books, understand from such sentences?
For example, in a review
in ``Le Monde" of Prigogine's most recent book \cite{P4},
Roger-Pol Droit notes that: ``The main discovery explained
in ``La fin des certitudes"
is the possibility to consider trajectories as probabilistic quantities
and to express the laws of dynamics in terms of ensembles." (\cite {Dro}).
If one understand probabilities in the classical sense, this is
perfectly acceptable, but it is not exactly a discovery (statistical
mechanics is more than a century old). And if it is
a major discovery, what are these
new laws of dynamics?}.
Either the dynamics is expressed directly at the level of
probability distributions, and we run into the difficulties mentioned in the
previous paragraph, or the dynamics is {\it fundamentally} expressed
in terms of trajectories (remembering that the discussion takes place
 in a classical
framework),  probabilities are a very useful tool, whose properties
are {\it derived} mathematically from those of the trajectories,
 and nothing radically new
has been done. In \cite{P2}\footnote{And in a private communication.},
Prigogine emphasizes  the ``irreducible" spectral
decompositions of the so-called Perron-Frobenius operator. This
is a rather technical notion, which I will discuss
in Appendix 2. It suffices  to say here that this will not solve the
dilemma raised above. If one reformulates the laws
of physics, or understands them differently, or whatever, there is still
presumably something that evolves, in some fashion. The question is:
what evolves, and how?

What the example of the billiard ball also
shows is that we must distinguish
different levels of analysis. First of all, we may describe the system
in a certain way: we may assign to the ball at least an approximate
position at each time, hence an approximate
trajectory\footnote{I say ``approximate", because I describe the
system as  it is seen. I do not yet consider
any theory (classical or quantum).}. Certainly the ball
is not {\it everywhere}, as the ``irreducible" probabilistic description would
suggest. The next thing we can do is to try to find exact or approximate
laws of motion for the ball. The laws of elastic reflection against obstacles,
for example. Finally, we may try to solve the equations of motion.
We may not be able to perform the last step. But this does not mean that one
should give up the previous ones.
 We may even realize that our laws are only approximate (because of friction,
of external perturbations, etc\dots). But why give up the
notion of (approximate) trajectories?
Of course, since we are not able to predict the evolution
of trajectories one may {\it choose} to study instead the evolution of probability
distributions. This is perfectly reasonable,
as long as one does not forget that, in doing so, we are   not
only studying the physical system but also our ability or inability to analyse it in
more detail. This will be very important in the next Section.

At this point, I want to briefly discuss the classical
status of probability in physics, i.e. of probability as ``ignorance".
This will also be very important in the next Section.
 To quote Laplace again: ``The curve described by a molecule of air
or of vapour is following a rule as certainly as the
orbits of the planets: the only
difference between the two is due to our ignorance.
Probabilility is related, in part to this ignorance, in part to our
knowledge."\cite{La} Let us consider the usual coin-throwing experiment.
We assign a probability $1/2$ to heads and $1/2$ to tails.
What is the logic of the argument?
We examine the coin, and  we find out that it is fair. We
also know the person who throws the coin and we know that he does not cheat.
But we are unable to control or to know exactly the initial conditions
for each throw. We can however determine the average result of a large
number of throws. This is simply because, if one considers
as a single ``experiment"  $N$ consecutive throws of a coin,
 the overwhelming majority (for $N$ large)
of the possible results will have an
approximately equal number of heads and
of tails. It is as simple as that, and there will be
nothing conceptually more
subtle in the way we shall use probabilities below.
 The part ``due to our ignorance" is simply that we
{\it use} probabilistic reasoning. If we were omniscient, it would not be needed
(but the averages would remain what they are, of course).
The part ``due to our knowledge" is what makes the reasoning work.
We could make a mistake: the coin could be biased, and we did not notice it.
Or we could have a ``record of bad luck" and have many more heads than tails.
But that is the way things are: our knowledge {\it is} incomplete and
we have to live with that. Nevertheless, probabilistic reasoning is
extraordinarily successful in practice, but, when it works, this is due
to our (partial) knowledge. It would be wrong to attribute any constructive
role to our ignorance. And it is also
erroneous to assume that the system must be somehow
indeterminate, when we apply probabilistic reasoning to it.
Finally, one could rephrase
Laplace's statement more carefully as
follows: ``Even if the curve described by a molecule of air
 follows a rule as certainly as the
orbits of the planets, our ignorance would force us to
use probabilistic
reasonings".

 \section{Irreversibility and the arrow of
time}
\begin{flushleft}
{\footnotesize Since in the differential equations of mechanics\\
themselves there is absolutely nothing analogous to the \\
Second Law of thermodynamics the latter can be mechanically \\
represented only by means of assumptions regarding initial conditions.\\
L. Boltzmann (\cite{Bo}, p.170)} \end{flushleft}

\vs{3mm}

\no {\bf 3.1. The problem.}

\vs{3mm}

What is the problem of irreversibility? The basic physical laws are
reversible, which simply means that, if we consider an isolated system of
particles, let it evolve for a time $t$, then reverse exactly the velocities
of all the particles, and let the system again evolve for a time $t$, we get
the original system at the initial time with all
velocities reversed\footnote{Mathematically, the microscopic state of the
system is represented by a point in its
 ``phase
space" $\Omega$.
Each point in that space
represents the positions and the velocities of {\it all} the particles
of the system under consideration. So the phase space is ${\bf R}^{6.N}$
where $N$ is the number of particles
(of the order of $10^{23}$ for a macroscopic system), since
one needs three coordinates for each position and three coordinates for
each velocity.  Hamilton's equations of motion
determine, for each time $t$, a map
 $T^t$ that associates to each initial
condition ${\bf x}\in \Omega$, at time zero,
the corresponding
solution
 $T^t {\bf x}$ of the equations of motion
at that time.
Reversibility of the  equations of motion means that
there is a transformation (an involution) $I$ acting on $\Omega$
 that satisfies the following relation:
\be
T^t I T^t {\bf x}=I{\bf x},
\en
or $IT^t=T^{-t}I$. In classical mechanics, $I$ reverses
velocities, without changing the positions. (In  quantum
mechanics, $\Omega$ is replaced by a Hilbert space, and
  $I$ associates to a wave function its complex conjugate.
For the role of weak interactions, see Feynman \cite{Fe2}.)
\protect\label{note_involution}}.
Now, there are lots of
motions that we see, without ever observing their associated
``time-reversed" motion: we go from life to death but not vice versa, coffee
does not jump out of the cup, mixtures of liquids do
not spontaneously unmix
themselves. Some of these  examples taken from
everyday life involve non-isolated
systems, but that is not relevant\footnote{We shall discuss
in Section 4.3  a
frequent confusion that assigns the {\it source} of irreversibility to the (true but
irrelevant) fact that no system is ever perfectly isolated. But let us
point out here that it is easy to produce non-isolated systems that behave
approximately in ``time reversed" fashion: a
refrigerator, for example. Living beings also seem to
violate the Second Law of
thermodynamics. But put a cat in a well-sealed box for a sufficiently long time
and
it will evolve towards equilibrium.}.
I shall center the discussion below
on the canonical physical example (and argue that the other
situations can be treated similarly): consider a gas that is
initially compressed by a piston in the left half of a box;
the piston is then released
so that the gas expands into the whole container.
We do not expect to see the
particles to go back to the left half of the box, although such a motion
would be as compatible with the laws of physics as the motion that
does take place.
So, the question, roughly speaking,
is this: if the basic laws are reversible, why do we see
 some motions but
never their time-reversed ones?

The first point to clarify is that this irreversibility does not lead
 to a {\it contradiction} with the basic physical laws\footnote{
Such a contradiction is suggested by the
following statement of Prigogine and Stengers: ``Irreversibility
is either true on all levels or on none: it cannot emerge as if out
of nothing, on going from one level to
another"(\cite{PS3}, quoted by Coveney, \cite{Co}, p.412.) Also,
``Irreversibility is conceivable only if the notion of point or of trajectory
loose their meaning" \cite{Cer}, p.166.}.
Indeed, the laws of physics are always of the
form: given some initial conditions, here is the result after some time.
But they never tell us how the world {\it
is or evolves}. In order to account for that, one
always needs to assume something about the initial conditions. The laws of
physics are compatible with lots of possible worlds: there could be no
earth, no life, no humans. Nothing of that would contradict the fundamental
physical laws. So, it is hard to see what kind of argument
 would imply  a
 contradiction between the reversibility of the laws and the
existence of irreversible phenomena. But no argument at all is given, beyond a
vague appeal to intuition, as for example: ``No speculation, no body of
knowledge ever claimed the equivalence between doing and undoing, between a
plant that grows, has flowers and dies, and a plant that resuscitates,
becomes younger and goes back to its primitive seed, between a man who
learns and becomes mature and a man who becomes progressively a child,
then an embryo
, and finally a cell. Yet, since its origins, dynamics, the physical theory
that identifies itself with the triumph of science, implied this radical
negation of time." (\cite{PS2}, p.25. The first of these sentences is
quoted again in \cite{P4}, p.178). But nobody says
that there is an ``equivalence" between the
two motions, only that both are compatible with the laws of physics. Which one,
if any, occurs depends on the initial conditions. And, if
the laws are deterministic, assumptions about the initial conditions
 are ultimately assumptions about the initial state
of the Universe.

 Once one has remarked
that, a priori, there is no contradiction
between irreversibility and the fundamental
laws, one could stop the discussion. It all depends on the
initial conditions, period. But this is rather unsatisfactory,
because, if one thinks about it, one realizes
that too many things could be ``explained"
by simply appealing to initial conditions. Luckily, much more can
be said. It is perfectly possible to give a natural
account of irreversible phenomena on
the basis of reversible fundamental laws, and of
suitable assumptions about initial
conditions. This was essentially done a
century ago by Boltzmann, and despite
numerous misunderstandings and misguided
objections (some of them coming from
famous scientists, such as Zermelo or Poincar\'e), his explanation still holds
today. Yet, Prigogine writes (\cite{P2}, p.41): ``He (Boltzmann) was forced
to conclude that the irreversibility postulated by thermodynamics was
incompatible with the reversible laws of
dynamics"\footnote{I. Stengers goes even further:``The reduction
of the thermodynamic entropy
to a dynamical interpretation can hardly be viewed otherwise than as an
``ideological claim"\dots" (\cite{St}, p.192). We shall see
below in what precise
sense this ``reduction" is actually a ``scientific claim".
}. This is in rather sharp
contrastwith Boltzmann's own words: ``From the fact that the
differential equations of mechanics are left unchanged by reversing the sign
of time without anything else, Herr Ostwald concludes that the mechanical
view of the world cannot explain why natural processes run preferentially in
a definite direction. But such a view appears to me to {\it overlook that
mechanical events are determined not only by differential equations, but also
by initial conditions}. In direct contrast to Herr Ostwald I have called it
one of the most brilliant confirmations of the mechanical view of Nature that
it provides an extraordinarily good picture of the dissipation of energy, as
long as one assumes that the world began in an initial state satisfying
certain initial conditions" (italics are mine; quoted in \cite{Le2},
replies, p.115).
I will now explain this ``brilliant confirmation
of the mechanical view of Nature", and show that all the alleged
contradictions are illusory\footnote{This is of course
not new at all. Good references, apart from Boltzmann
himself, \cite{Bo}, include Feynman \cite{Fe2},
Jaynes \cite{Ja}, Lebowitz \cite{Le1,Le2},
Penrose \cite{Pe2}, and Schr\"odinger
\cite{Sc}.}.

\vs{3mm}

\no {\bf 3.2. The classical solution.}\footnote{By classical,
I mean ```standard". However, all the discussion will take place in the context
of classical physics. I will leave out quantum mechanics entirely.
Although the quantum picture may be more complicated, I do not believe that
it renders obsolete the basic ideas explained here.}

\vs{3mm}

First of all, I should say that Boltzmann gives a {\it framework}
in which to account for irreversible phenomena on the basis
of reversible microscopic laws. He does not explain in detail
every concrete
irreversible phenomenon (like diffusion, or the growth of a plant). For that,
more work is needed and, while the general framework that I shall
discuss uses very little of the  properties of the microscopic
dynamics, the latter may be important in the explanation
of specific irreversible phenomena\footnote{This is analogous to the
theory of natural selection. The latter provides a scheme
of explanation for the appearance of complex organs,
but more detailed arguments are needed
to account for concrete properties of living beings. See Section 6
for further discussion of this analogy.}.

Let
us now
see which systems do behave irreversibly. A good test is to record the
behaviour of the system in a movie, and then to run the movie backwards. If it
looks funny (e.g. people jump out of their graves), then we are facing
irreversible behaviour. It is easy to convince oneself that all the
familiar examples
of irreversible behaviour involve systems with a large number of
particles (or  degrees of
freedom). If one were to make a movie of the motion of one molecule, the
backward movie would look completely natural. The same is true for a
billiard ball on a frictionless billiard table\footnote{Of course, the billiard
ball itself contains many molecules. But the
rigidity of the ball allows us to concentrate on the motion of its center of
mass.}. If, however, friction is present, then we are dealing with
many degrees of freedom (the atoms in the billiard table, those in the
surrounding air etc...).

There are two fundamental ingredients in the
classical explanation of irreversibility, in addition to the microscopic
laws.
The first has  already been introduced: initial conditions.
The second  is suggested by the remark
that we deal with  systems with many degrees of freedom: we {\it have} to
distinguish between microscopic and macroscopic variables.
 Let us consider the phase space $\Omega$
(see note \protect\ref{note_involution})
of
the system, so that the system is represented by a point ${\bf x}$ in that
space and its  evolution is represented by a curve
${\bf x}(t)=T^t({\bf x})$. Various quantities of physical interest,
for example the density, or the average energy, or the average velocity in a
given cubic millimeter, can be expressed as functions on
$\Omega$\footnote{By ``functions", I mean also families of functions
indexed by space or time, i.e.
fields, such as the local energy density, or the velocity field.}.
These functions (call  them $F$)
tend to be many-to-one, i.e. there are typically
a huge number of configurations giving rise to a given
value of $F$\footnote{I am a little vague on how to ``count" configurations.
If I consider discrete (finite) systems, then it is just counting.
Otherwise, I use of course the Lebesgue measure on phase space.
All statements about probabilities made later will be
based on such ``counting".}. For
example, if $F$ is the total energy, then it
takes a constant value on a surface in phase space. But if $F$
is the number of particles in a cubic millimeter, there are also
lots of microscopic configurations corresponding to a given value of $F$.
Now, let me make two statements, the
first of which is trivial and the second  not. Given a
microscopic initial configuration ${\bf x}_0$, giving rise to a trajectory
${\bf x}(t)$, any function on phase space follows an induced evolution
$F_0 \rightarrow F_t$, where $F_0 =F({\bf x}_0)$, and
$F_t=F({\bf x}(t))$ (here and below, I shall assume that $t$
is positive).
That is the trivial part. The non-trivial
observation is that, in many situations, one can find a suitable
family of functions (I'll still denote by $F$ such a family)
 so that this
induced evolution is actually  (approximately)
$autonomous$. That is, one can determine
$F_t$ given $F_0$ alone, without having to know
the microscopic configuration
from which it comes\footnote{Although not trivial,
this expresses the fact  that reproducible macroscopic experiments
exist and that a deterministic macroscopic
description of the world is possible.}. This means that the
different microscopic
configurations on which $F$ takes the value $F_0$, will induce the same
evolution on $F_t$. A very trivial example is given by the
globally conserved quantities (like the total energy):
for all microscopic configurations, $F_t = F_0$, for all times. But that is
not interesting. It is more interesting to observe that the solutions
of all the familiar
macroscopic equations (Navier-Stokes, Boltzmann, diffusion, \dots) can be
considered as defining such an induced evolution $F_0 \rightarrow F_t$.
 Actually, there are several provisos to be made
here: first of all, it is not true
that {\it all}
microscopic configurations giving rise to $F_0$ lead to the same evolution
for $F_t$. In general,  only the (vast) majority of microscopic
configurations  do that\footnote{To make the
micro/macro distinction sharp, one has to consider
some kind of
limit (hydrodynamic, kinetic, etc\dots), where
the number of particles (and other quantities)
tend to infinity.
That is a convenient mathematical setting to prove precise statements.
But one should not confuse this limit, which is an approximation
to the real world,
with the physical
basis of irreversibility. See Lebowitz \cite{Le1} and Spohn \cite{Sp}
for a discussion of those
limits.}. Moreover, if we want that evolution to
hold for all times,
then this set of microscopic configurations may
become empty\footnote{This is due, for example, to the Poincar\'e recurrences, see
Section 4.1 and Appendix 1.}. Finally, the laws used in practice  may contain
some further  approximations.

So, the precise justification of a macroscopic law should be
given along the following lines: given $F_0$, and
given a (not too large) time $T$\footnote{I mean shorter than
the Poincar\'e recurrence time.}, there exists a large subset of the set of
${\bf x}$'s giving rise to $F_0$ (i.e. of the preimage in $\Omega$, under
the map $F$, of $F_0$) such that the
induced evolution of $F_t$ is approximately
described by the relevant macroscopic equations up to time $T$. It
should be obvious that it is not easy  to prove  such a
statement. One has to deal with dynamical systems with a large number of
degrees of freedom, about which very little is known, and in addition
one has to
identify limits in which one can make sense of the approximations mentioned
above (a large subset, a not too large time $T$ \dots). Nevertheless, this
can be done in some circumstances, the best known being probably the
derivation of Boltzmann's equation by Lanford \cite{La1,La2,La3}. In
 Appendix 1, I
discuss a model due to Mark Kac which, while artificially simple, can be
easily analysed and shows exactly what one would like to do in more
complicated situations\footnote{It is also important to clarify
the role of ``ensembles" here. What we have to explain is the fact that,
when a system satifies certain macroscopic initial conditions ($F_0$)
 it {\it always} (in practice)
obeys certain macroscopic laws. The same macroscopic initial conditions
will correspond to many different microscopic initial conditions. We may
introduce, for mathematical convenience, a probability distribution
(an ``ensemble") on the microscopic initial conditions. But one should
remember that we are physically
interested in ``probability one statements", namely
statements that hold for (almost) all
 microscopic configurations, as opposed
to statements about averages, for example.
Otherwise, one would not explain why all individual macroscopic systems
satisfy a given law. In practice, those ``probability one statements"
will only hold in some limit. Physically, they should be
interpreted as ``very close to one" for  finite systems with a large number
of particles. To see how close to one
this is, consider all the bets and games of chance
that have ever taken place in human history. One would certainly expect
the laws of large numbers to apply to such a sample. But this number
is minuscule compared to the typical number ($10^{23}$) of molecules
in a cubic centimeter.
\protect\label{note_numbers}}.

Let us come back to the problem of irreversibility: should we expect those
macroscopic laws to be reversible? A priori, not at all.
Indeed, I have emphasized in the abstract description above the role of
initial conditions in their derivation\footnote{This
remark is of some interest
for the issue of {\it reductionism}: higher level laws, such as the
macroscopic laws, are reduced to the microscopic ones {\it plus}
assumptions on the
initial conditions. If this is  the case in statistical mechanics,
where it is usually granted that reductionism works, it should
clarify the situation in other fields, like biology, where reductionism
is sometimes questioned. In particular, the fact that some assumptions
must be made on the initial conditions in going from the
microscopic to the macroscopic should not be forgotten, nor should it
be held as an argument against reductionism. Another frequent confusion
about reductionism is to remark that the macroscopic laws do not
uniquely determine the microscopic ones. For example, many of
the macroscopic laws can be derived  from stochastic
microscopic laws or from deterministic ones.
That is true, but does not invalidate
reductionism. What is true on the microscopic level has to be discovered
independently of the reductionist programme. Finally, what is considered
microscopic or macroscopic is a question of scale. The classical description
considered here at the ``microscopic" level is an approximation to the
quantum description and neglects the molecular and atomic structure.
And the ``macroscopic" level may in turn be considered microscopic
if one studies large-scale motions of the atmosphere. But, despite frequent
claims to the contrary, reductionists are quite happy not to explain
carburetors directly in terms of quarks (see Weinberg
\cite{We} for a good
discussion of reductionism).}.
 The macroscopic equations may be reversible or not, depending
on the situation. But since {\it initial} conditions enter
their derivation, there is no
{\it logical} reason to expect them to be reversible\footnote{Note
that I am not discussing irreversibility in terms of the increase of
entropy, but
rather in terms of the macroscopic laws. After all, when
we observe the mixing of different fluids, we see a phenomenon described
by the diffusion equation, but we do not see entropy flowing. The connection
with entropy will be made in Section 5.}.

\vs{3mm}

\no {\bf 3.3. The reversibility objection.}

\vs{3mm}

Let me illustrate this explanation of irreversibility
in a concrete physical example (see
also Appendix 1 for a simple mathematical model). Consider the gas
introduced in Section 3.1 that is
initially compressed by a piston in the left half of a box, and that
expands  into the whole box. Let $F$ be the density
of the gas. Initially, it is one (say) in one half of the box and zero in the
other half. After some time $t$,
 it
is (approximately)
one half everywhere. The explanation of the
irreversible evolution of $F$ is
 that the overwhelming majority of the microscopic configurations
corresponding to the gas in the left half, will evolve deterministically
so as to induce the observed evolution of $F$. There may of course be some
exceptional configurations, for which all the particles stay in the left half.
All one is saying is that those configurations are extraordinarily rare,
and that we do not expect to see even one of them appearing when
we repeat the experiment many times, not even
once ``in a million years", to put
it mildly \cite{Fe2} (see the end of note
(\protect\ref{note_numbers})).

This example also illustrates the answer to the reversibility objection.
Call ``good"  the microscopic configurations that lead to the
expected macroscopic behaviour. Take all the good microscopic
configurations in the left half of the box, and let them evolve until
the density is approximately uniform.
Now, reverse all the velocities. We get a set of configurations that still
determines a density one half in the box. However, they are not good.
Indeed, from now on, if the system remains isolated, the density just remains
uniform according to the macroscopic laws. But for the
 configurations just described, the gas will move back
to the left half, leading to a gross violation of the macroscopic law. What is
the solution? Simply that those ``reversed-velocities" configurations form
a very tiny subset of all the microscopic configurations giving rise to a
uniform density. And, of course, the original set of configurations, those
coming from the left half of the box, also form such a small subset.
Most configurations corresponding to a uniform density
do not go to the left half of the box, neither in the future nor in the past
(at least for reasonable periods of time, see Sect. 4.1).
 So that, if we prepare the system with a uniform density,
we do not expect to ``hit" even once one of those bad
 configurations\footnote{To put it
in formulas,
let  $\overline \Omega_t $ be
the configurations giving to $F$ its value at time $t$.
If we denote by $F_t$ that value,  $\overline \Omega_t $ is
simply the preimage of $F_t$ under the map $F$.
Let
 $\Omega_t$ be the set of good configurations, at
time $t$, that  lead to a
behaviour of $F$, for later times (again, not for {\it too} long, because
of Poincar\'e recurrences), which is  described by the macroscopic
laws. In general, $\Omega_t$ is a  very large
subset of $\overline \Omega_t$, but is not identical to $\overline \Omega_t$.
Thus, $\overline \Omega_0$ are all the
configurations
in the left half of the box at time zero, and
 $\Omega_0 $ is the subset consisting of those configurations
whose evolution lead to a uniform density.
Microscopic reversibility says that $T^t
(I(T^t (\Omega_0)))=I(\Omega_0)$ (this is just (1) in note
(\protect\ref{note_involution})
applied to $\Omega_0$).
A reversibility paradox
would follow from $T^t(I(\Omega_t))=I(\Omega_0)$
(one takes all the good configurations at time $t$, reverses
their velocities, lets
them evolve for a time $t$ and thereby gets the original set of initial conditions,
with velocities reversed). But $\Omega_t$ is {\it not equal}, in general,
 to $T^t
(\Omega_0)$ (and this is the source of much confusion).
 In our  example, $T^t
(\Omega_0)$ is a tiny subset of $ \Omega_t$, because most configurations
in $\Omega_t$ were not in the left half of
the box at time zero. Actually, $I(T^t (\Omega_0))$ provides an example
of configurations that belong to $\overline \Omega_t$ but not to $ \Omega_t$.
These configurations
 correspond to a uniform density at time $t$, but not at time $2t$.
\protect\label{note_good}}.

Now comes a real problem. We are explaining that we never expect to get
a microscopic configuration that will lead all the gas to the left of the box.
{\it But we started from such a configuration}. How did we get there in the
first place? The real problem is not to explain why one goes to equilibrium,
but why there are systems out of equilibrium to start with.
For the gas, obviously the system was not isolated: an experimentalist
pushed the piston. But why was there an experimentalist? Human beings
are also systems out of equilibrium, and they remain so (for some time)
 thanks to the food they eat, which itself depends on
the sun, through the plants and their photosynthesis. Of course,
in order to be able to take advantage of their food,
 humans also need their genetic program, which itself
results from the long history of natural selection.

 As discussed e.g. in Penrose \cite{Pe2}, the earth
does not gain energy from the sun
(that energy is re-radiated by the earth), but
low entropy (likewise, we seek low entropy rather than energy
 in our food); the sun sends (relatively) few high energy photons
and the earth re-radiates more low energy photons (in such a way that the
total energy is conserved). Expressed in terms of ``phase space",
 the numerous low energy photons occupy a much bigger volume
than the incoming high energy ones. So, the
solar system, as a whole, moves towards a larger part of its phase space
 while the sun burns its fuel.
That evolution accounts, by far, for
what we observe in living beings or in other ``self-organized"
structures\footnote{Failure to realize this leads
to strange statements, as for example in
Cohen and Stewart (\cite{CS}, p.259): speaking of the evolution since
the Big Bang, the authors write:
``For systems such as these, the thermodynamic model of independent
subsystems whose interactions switch on but not off is simply irrelevant.
The features of thermodynamics either don't apply or are so long term that
they don't model anything  interesting. Take Cairns-Smith's scenario
of clay as scaffolding for life. The system consisting of
clay alone is {\it less}
ordered than that of clay plus organic molecules: Order
is increasing with time. Why?"
The explanation given afterwards
 ignores both the action of the sun, and the original
``improbable state" discussed here. As Ruelle wrote in a review of
this book  ``if life violates the Second Law,
why can't one build a power plant (with some
suitable life forms in it) producing ice cubes and water currents
from the waters of Loch Ness?" (\cite{Ru2}).
 A similar confusion can be
found in Popper: ``This law of the increase of disorder, interpreted as a
cosmic principle, made the evolution of life incomprehensible, apparently
even paradoxical." (\cite{Pop5}, p.172)\protect\label{note_life}}.
I shall come back to this point in Section 6.
Of course, for the sun to play this role, it has to be itself
out of equilibrium, and to have been even more so in the past.
We end up with an egg and hen problem and we have
ultimately to assume that the Universe started in a
 state far from equilibrium,
an ``improbable state" as Boltzmann called it.
To make the analogy with the gas in the box,
it is as if the Universe had started in a very
little corner of a huge box\footnote{I neglect
here the effect of gravity: see Penrose \cite{Pe2}.}.

To account in
a natural way for such a state is of course a major open problem, on which
I have nothing to say (see Penrose
\cite{Pe2} for further discussion,
and
Figure 7.19 there for an illustration),
 except that one cannot avoid it by ``alternative"
explanations of irreversibility. Given the laws of physics, as
they are formulated now, the world could have started in
equilibrium,
and then we would not be around to discuss
the problem\footnote{As Feynman says: ``Therefore
I think it necessary to add to the physical laws
the hypothesis that in thepast the universe was more ordered, in the
technical sense, than it is today - I think this is the additional
statement that is needed to make sense, and
to make an understanding of the
irreversibility." (\cite{Fe2}, p.116)\protect\label{note_Feynman}}.
To summarize: the
only real problem with irreversibility is not
to explain irreversible behaviour in the future,
but to account for the ``exceptional"
 conditions of the Universe in the past.

\vs{3mm}

\no {\bf 3.4. Chaos and irreversibility.}

\vs{3mm}

Now, I come to my basic criticism of the
views of Prigogine and his collaborators, who argue that dynamical systems with very
good chaotic properties, such as the baker's map, are ``intrinsically
irreversible".
Let me quote from a letter of a collaborator of Prigogine,
D. Driebe \cite{Dr},
 criticizing  an
article of Lebowitz
\cite{Le2} explaining Boltzmann's ideas. This letter is
remarkably clear and summarizes well the main points of disagreement.
 ``If the scale-separation argument were the whole story,
then irreversibility would be due to our approximate observation or
limited knowledge of the system. This is difficult to reconcile
with the constructive role of irreversible processes\dots Irreversibility
is not to be found on the level
 of trajectories or wavefunctions but is instead manifest on the level
of probability distributions\dots Irreversible processes are well observed in
systems with few degrees of freedom, such as the baker and the
multibaker transformations\dots The arrow of time is not due to some
phenomenological approximations but is an intrinsic property of classes
of unstable dynamical systems"\footnote{In a recent textbook one reads,
after a
discussion of the baker's map: ``Irreversibility appears only because the
instantaneous state of the system cannot be known with an infinite
precision" (\cite{Va}, p.198).}.

Let us discuss these claims one by one. First of all,
as I emphasized above, the scale-separation  (i.e. the
micro/macro distinction) is not ``the whole story". Initial conditions
have to enter into the
explanation (and also the dynamics, of course). Next,
what does
it mean that ``irreversible processes are observed in
systems  such as the baker transformation"? This transformation
 describes  a chaotic system with few degrees of freedom,
somewhat like the billiard ball on a frictionless table\footnote{The baker
map is
quite similar to the map discussed in note
\protect\ref{note_10x}, and has the same chaotic
properties as the latter, but is invertible.}.
For those systems, there is no sense of a micro/macro distinction:
how could one define the macroscopic variables?
To put it otherwise, we can make a movie of the
motion of a point in the plane evolving under the baker's map, or of a
billiard ball, or of any isolated chaotic system with few degrees of
freedom, and run it backwards, we shall not be able to tell the difference.
There is nothing funny or implausible going on,
unlike the backward movie of any real irreversible
macroscopic phenomenon.
So, the first critique of this alleged connection between unstable
 dynamical systems (i.e. what I call here chaotic systems) and irreversibility
is that one ``explains" irreversibility in systems in which nothing
irreversible happens, and where  therefore there is nothing  to be explained.

 It is true that probability distributions for those systems evolve
``irreversibly", meaning that any (absolutely continuous, see
note \protect\ref{note_proba})
probability distribution will spread
out all over the phase space and will quickly tend to a uniform distribution.
This just reflects the fact that different points in the support of the
initial distribution, even if they are close to each other initially, will
be separated by the chaotic dynamics. So, it is true,
in a narrow sense, that
``irreversibility
 is  manifest on the level
of probability distributions". But what is the physical meaning of this
statement? A physical system, chaotic or not, is described by a
trajectory in phase space,
and is certainly not described adequately
by the corresponding probability distributions. As I discussed in Section
2.2, the latter
 reflects, in part, our ignorance of that trajectory. Their ``irreversible"
 behaviour in this sense is therefore not a genuine physical
property of the system. We can, if we want, focus our attention
on probabilities rather than on trajectories, but that ``choice" cannot
have a basic role in our explanations.

One cannot stress strongly enough the difference between the role played
by probabilities here and in the classical solution. In the latter, we
use probabilities as in the coin-throwing experiment. We
have some macroscopic
constraint on a system (the coin is fair; the particles
are in the left half of the box),
corresponding to a variety of microscopic configurations.
We predict that the behaviour of certain macroscopic
variables (the average number of heads; the average density)
will be the one induced by the vast majority of microscopic configurations,
compatible with the initial constraints. That's all. But it works
only because a large number of variables are involved, {\it in each single
physical system}. However, each such system is described by a point
in phase space (likewise,
the result of many coin throwings is a particular
sequence of heads and tails).
In the ``intrinsic
irreversibility" approach, a probability distribution is assigned to
{\it each single
physical system}, as an ``irreducible" description. The only way
I can make sense of that  approach is to
consider a {\it large number} of billiard balls or of copies of the
baker's map, all of them starting with nearby initial conditions.
Then, it would be like the particles in the box,  the average density would
tend to become uniform, and we are back to the standard picture.
But this does not force us to ``rethink the notion of law of nature".

\vs{3mm}

\no {\bf 3.5. Is irreversibility subjective?}

\vs{3mm}

I will now discuss the alleged ``subjectivity" of
this account of irreversibility
(i.e., that it is due to our approximate observation or
limited knowledge of the system). I shall consider
in Section 6 the ``constructive role"
of irreversible processes,
mentioned in Driebe's letter \cite{Dr}.
Branding Boltzmann's ideas as ``subjective"
is rather common. For example, Prigogine writes:
``In the classical picture, irreversibility was due to our
approximations, to our ignorance."
(\cite{P2}, p.37) But,
thanks to the existence of
unstable dynamical systems, ``the notion of probability that
Boltzmann had introduced in order to express the
arrow of time does not correspond to our
ignorance and acquires an objective
meaning" (\cite{P2}, p.42)\footnote{See e.g. Coveney (\cite{Co}, p.412):
 ``Another
quite popular
approach has been to relegate the whole question of irreversibility
as illusory." See also Lestienne (\cite{Les}, p.176) and
Prigogine and Stengers (\cite{PS1} p.284) for similar remarks.}.
To use Popper's image: ``Hiroshima is not an
illusion" (I shall come back to Popper's confusions in Section 4.4.).
 This is only a dramatization of the fact that irreversible
events are not subjective, or so it seems. The objection is that, if
the microscopic variables behave reversibly and if irreversibility only
follows when we {\it  ``choose"} to concentrate our attention on macroscopic
variables, then our explanation of irreversibility is unavoidably tainted by
subjectivism. I think that this charge is completely unfair, and reflects
some misunderstanding of what irreversible phenomena really are. The point
is that, upon reflection, one sees that all irreversible phenomena  deal
with these macroscopic variables. There is no subjectivism here: the
evolution of the macroscopic variables is objectively determined by
the microscopic ones, and they behave as they do whether we look at them or not.
In that sense they are completely objective. But it is true that, if we look at
a single molecule, or at a collection of molecules represented by a point in
phase space, there is no sense in which they  evolve ``irreversibly", if we
are not willing toconsider some of the macroscopic variables that they
determine.

However,  the
apparently ``subjective" aspect of irreversibility
has been sometimes overemphasized, at least as a way to speak. Heisenberg
wrote: ``Gibbs was the first to introduce
a physical concept which can only be applied to an
object when our knowledge of the object is
incomplete. If for instance the motion and the
position of each  molecule in a gas were
known, then it would be pointless to
continue speaking of
the temperature of the gas."(\cite{He}, p.38)\footnote{Pauli made a
similar remark, see \cite{Pa}, quoted in Popper,
\cite{Pop4}, p.109.}. And Max Born said:
``Irreversibility is therefore a consequence of the explicit introduction of
ignorance into the fundamental laws." (\cite{Bo2}, p.72). These formulations,
although correct if they are properly interpreted, lead to
unnecessary confusions.
For example, Popper wrote: ``It is clearly
absurd to believe that pennies fall
or molecules collide in a random fashion
{\it because we do not know} the initial
conditions, and that they would do otherwise if some demon were to give
their secret away to us: it is not only impossible, it is absurd to explain
objective statistical frequencies by
 subjective ignorance." (\cite{Pop4}, p.106)\footnote{In his textbook on
Statistical Mechanics; S.-K. Ma shows similar concerns:
``In one point of view,
 probability expresses the knowledge of the observer.
If he knows more about the
system, the probability is more concentrated. This is obviously incorrect.
The motion of the system is independent of the psychological condition of the
observer." (\cite{Ma1}, p.448) And H. Bondi wrote: ``It is somewhat offensive
to our thought to suggest that if we know a system in detail then
we cannot tell which way time is going, but if we take a blurred view, a
statistical view of it, that is to say
throw away some information, then we can\dots" (\cite{Bon},
quoted in \cite{Lan}, p.135. T. Gold expressed similar
views, see \cite{Lan}).}.
However, just after saying this,
Popper gives what he calls ``an objective probabilistic explanation
of irreversible processes" (\cite{Pop4}, p.107), attributed
to Planck, which, as far as I can tell,
is not very different from what I call the classical solution.
The source of the confusion comes from two uses of the word ``knowledge".
Obviously, the world does what it does, whether we know about it or not.
So, indeed, if ``some demon" were to provide us with a detailed knowledge
of the microscopic state of the gas in the left half of the box, nothing
would change to the future evolution of that gas. But we may imagine situations
where one can {\it control} more variables, hence to ``know" more about the system.
When the piston forces the gas to be in the left half of the box,
the set of available microscopic states is different
than when the piston is not there, and obviously we have to take that
``knowledge" into account. But there is nothing mysterious here.

I believe
that statistical mechanics would become  easier to
understand by students if it were presented
without using an anthropomorphic language and
subjective sounding notions such as
information, observation or knowledge. Or, at least, one should
explain precisely why these notions
are introduced and why they do not contradict an objectivist
view of natural phenomena (see the writings of
Jaynes on this point \cite{Ja,Ja4}).
But I also believe that the charge of subjectivity should be
completely reversed:
to ``explain" irreversibility through the
behaviour of probability distributions
(which {\it are} describing our ignorance), as Prigogine does,
 is to proceed as if the
 limitations of human knowledge
played a fundamental physical role.

\section{Some misconceptions about irreversibility}
\begin{flushleft}
{\footnotesize The Second Law can never be proved mathematically \\
by means of the equations of dynamics alone.\\
L. Boltzmann (\cite{Bo}, p.204).} \end{flushleft}

\vs{3mm}

\no {\bf 4.1. The Poincar\'e recurrence theorem.}

\vs{3mm}

According to Prigogine (\cite{P2}, p.23) Poincar\'e did not recommend
reading Boltzmann, because his conclusions were in contradiction with his
premises. Discussing our example of a gas expanding in a container,
Prigogine observes that ``if irreversibility was only that,
it would indeed be an illusion, because, if we wait even longer, then it may
happen that the particles go back to the same half of the container. In this
view, irreversibility would simply be due to the limits of our patience."
(\cite{P2}, p.24) This is basically the argument derived from the Poincar\'e
recurrence theorem (and used by Zermelo
against Boltzmann \cite{Ze}), which says that, if the container remains isolated long
enough, then indeed the particles will return to the half of the box
from which they started. Replying to that argument, Boltzmann
supposedly said ``You
should live that long". For any realistic macroscopic
system, the Poincar\'e recurrence
times (i.e. the time needed for the particles to return to the left half of
the box) are much much larger than the age of the universe. So that again no
contradiction can be derived, from a physical point of view, between
Boltzmann's explanations and Poincar\'e's theorem. However, there is still a
mathematical problem (and this may be what Poincar\'e had in mind): if
one tries to rigorously derive an irreversible macroscopic equation from
the microscopic dynamics and suitable assumptions on initial conditions, the
Poincar\'e recurrence time will put a limit on the length of the
time interval over which
these statements can be proven. That is one of the reasons why one discusses
these derivations in suitable limits (e.g. when the number of particles goes
to infinity) where the Poincar\'e recurrence time becomes infinite. But one
should not confuse the fact that one takes a limit for mathematical
convenience and the source of irreversibility. In the Kac model discussed in
Appendix 1, one sees clearly that there are very different time scales:
one over which convergence to equilibrium occurs, and a much larger one,
where the Poincar\'e recurrence takes place. But the first time scale is not
an ``illusion". In fact, it is on that time scale that all phenomena that we
can possibly observe do take place.

\vs{3mm}

\no {\bf 4.2. Ergodicity and mixing.}

\vs{3mm}

One often hears that, for a
system to reach ``equilibrium", it must be ergodic, or mixing. The fact is that
those properties, like the ``intrinsic irreversibility"
discussed above, {\it are neither necessary nor  sufficient}
for a system to approach equilibrium. Let me start with ergodicity.
A dynamical system is {\it ergodic} if the average time
spent by a trajectory in any region of
the phase space is proportional to the volume of that region.
To be more precise: average means in the limit of infinite time  and this
property has to hold for all trajectories, except (possibly) those
lying in a subset of zero volume. One says that it holds for ``almost all"
trajectories. This property implies that, for any reasonable function on
phase space, the average along almost all
trajectories will equal the average over the phase space\footnote{In formulas,
let $\Omega$ be the ``phase space" on which the motion is ergodic (
 i.e. a constant energy
surface, which is a subset of the space considered in note
\protect\ref{note_involution},
on which is defined the measure induced by the Lebesgue measure, normalised
to one, and denoted $d{\bf x}$). Then,
ergodicity means that for $F$ integrable,
\be
\lim_{T \rightarrow \infty} \frac{1}{T} \int_0^T F(T^t {\bf x}) dt =
\int_{\Omega} F({\bf x}) d{\bf x},
\en
 for almost all initial conditions ${\bf x} \in \Omega$.
The LHS is the time average and the RHS the space average.
If we take for $F$ the characteristic function of a (measurable) set $A\subset \Omega$,
the time average equals the fraction of time spent by the trajectory in $A$, and
the space average is the volume of $A$.}.
Then, the argument goes, the measurement
of any physical quantity will take some time. This time is long
compared to the ``relaxation time"
of molecular processes. Hence, we can approximately regard it as infinite.
Therefore, the measured quantity, a
time average, will approximately equal the average over phase space of the physical quantity
under consideration. But this latter average is exactly what one
calls the equilibrium value
of the physical quantity. So, according to the usual
story,
if a dynamical system is ergodic, it converges
towards equilibrium. This appeal to ergodicity in order to justify
statistical mechanics is rather widespread\footnote{
For a history of the concept of ergodicity, and  some very
interesting
modern developments, see Gallavotti \cite{Gal}.
It seems that the (misleading) emphasis on the modern notion
of ergodicity goes back to the Ehrenfests' paper \cite{Eh},
more than to Boltzmann.
A careful, but
nevertheless exaggerated interest in ergodicity
and mixing is found in the work of Khinchin \cite{Kh} and Krylov \cite{Kr};
 it is also found e.g.
in Chandler \cite{Ch}, p.57,
Hill \cite{Hi}, p.16, S. K. Ma \cite{Ma1} Chap.26, Thompson
\cite{Tho}, App.B,
and Dunford and Schwartz \cite{DS}, p.657 (but see Schwartz \cite{Sch}
for a self-criticism of \cite{DS});
 in the recent textbook of Vauclair
\cite{Va}, one reads: ``One considers that
during the time $\delta t$ of the measurement, the system has gone through
all the possibly accessible states, and that it spent in each state
a time proportional to its probability." (p.11) And:  ``Only the
systems having this property (mixing)
 tend to an equilibrium state, when they are initially in a state
out of
equilibrium." (p.197).}
even though it has been properly
criticized  for a long time   by, e.g., Tolman
\cite{To}, p.65, Jaynes \cite{Ja}, p.106, and Schwartz \cite{Sch}.

Let us see the problems with this argument: a well-known, but
relatively minor, problem
is that it is very hard to give a mathematical proof that a realistic
mechanical system is  ergodic. But let us take such a proof for granted,
for the sake of the discussion. Here is a more serious problem. Assume that
the argument given above is true: how would it then be possible to observe or
measure {\it any non-equilibrium} phenomenon? In the experiment with the box
divided in two halves, we should not be able to see any intermediate stage,
when the empty half gets filled, since the time for our measurements is
supposed to be approximately infinite. So, where is the problem? We
implicitly identified the ``relaxation time" with what one might call the
``ergodic time", i.e. the time taken by the system to visit all regions of
phase space sufficiently often so that the replacement of time averages by
spatial averages is approximately true. But, whatever the exact meaning of
the word ``relaxation time" (for a few molecules) is, the ergodic time is
certainly enormously longer. Just consider how large is the volume in phase
space that has to be ``sampled" by the trajectory. For example,
 all the particles could be in the right half of the box, and ergodicity
says that they will spend some time there (note that this is not implied
by Poincar\'e's theorem; the latter only guarantees that the particles
will return to the part of the box from which they started, i.e. the left half here).
To be more precise, let
us partition the  phase space into a certain number of cells, of a
given volume, and consider the time it takes for a given trajectory to visit
each cell, even once, let us say\footnote{If there is some cell which has
not been visited even once, there will be a function on phase space for
which the space average and the time average, computed up to that time, differ
a lot: just take the function which takes value one on that cell and is zero
elsewhere.}. That, obviously, will depend on the size (hence, on the number)
of the cells. By taking finer and finer partitions, we can make that time as
large as one wishes. So, if one were to take the argument outlined
above literally, the ``ergodic time" is infinite, and speaking loosely about
a relaxation time  is simply misleading.

At this point of the discussion, one often says that we do not need the
time and space average to be (almost) equal for all functions, but only for
those of physical relevance (like the energy or particle densities). This
is correct, but the criticism of the ``ergodic" approach then changes:
instead of not being {\it sufficient} to account for irreversibility, we observe
that it is not {\it necessary}. To see this, consider another partition of phase
space: fix a set of macroscopic variables, and partition the phase space
according to the values taken by these variables (see e.g. figures 7.3
and 7.5 in Penrose
\cite{Pe2}  for an illustration, and
 Appendix 1 here for an example). Each element of the
partition consists of a set of microscopic states that give the same value
to the chosen macroscopic variables. Now, these elements of the partition
have very different volumes. This is
similar to  the law of large
numbers. There are (for $N$ large) vastly more results of $N$ throws of a
coin where the number of heads is approximately one half than throws where
it is approximately one quarter (the ratio of these two numbers varies
exponentially with $N$). By far the largest volumes correspond to the
{\it
equilibrium values} of the macroscopic variables (and that is how
``equilibrium" should be defined). So, we need a much weaker
notion than ergodicity. All we need is that the microscopic
configuration evolves in phase space towards those regions where the
relevant macroscopic variables take their equilibrium values. The Kac model
(see Appendix 1) perfectly illustrates  this point: it is not ergodic in
any sense, yet, on proper time scales, the macroscopic variables evolve
towards equilibrium.

There is a hierarchy of ``ergodic" properties that are stronger than
ergodicity: mixing, K-system, Bernoulli, see
Lebowitz and Penrose \cite{LP}. But none of these will
help us to understand, in principle,
 irreversible behaviour any more than ergodicity.

The problem with all those approaches is that they try to give a purely
mechanical criterion for ``irreversible behaviour". Here is the basic
dilemna: either we are willing to introduce a macro/micro distinction and to
give a basic role to initial conditions in our explanation of
irreversibility or we are not. If we make the first choice, then, as
explained in Section 3, there is no deep problem with irreversibility, and
subtle properties of the dynamics (like ergodic properties) play basically
no role. On the other hand, nobody has ever given a consistent alternative,
namely an explanation of irreversibility that would hold for {\it all} initial
conditions or apply to {\it all} functions on configuration space (therefore
avoiding the micro/macro distinction). So, we have to make the first
choice. But then, everything is clear and nothing else is needed.

Another critique of the ``ergodic" approach is that systems with one or few
degrees of freedom may very well be ergodic, or mixing, or Bernoulli
(like the baker's transformation). And,
as we discussed in Section 3.4, it makes no sense to speak about
irreversibility for those systems. So, this is another sense in which the
notion of ergodicity is not sufficient (see e.g. Vauclair
(\cite{Va} p.197), where the
approach to equilibrium is illustrated by the baker's transformation).

To avoid any misunderstandings, I emphasize that the study of ergodic
properties of dynamical systems gives us a lot of interesting information
on those systems, especially for chaotic systems.
Besides, ergodic properties, like other
concrete dynamical properties of a system,
may play a role in the form of the macroscopic equations
obeyed by the system, in the value of some transport coefficients
or in the speed of convergence to equilibrium.
But, and this is the only point I wanted to make, the usual
story linking
 ergodicity (or mixing) and ``approach to equilibrium"
is highly unsatisfactory.

 \vs{3mm}

\no {\bf 4.3. Real systems are never isolated.}

\vs{3mm}

Sometimes it is alleged that, for some reason (the Poincar\'e recurrences,
for example) a truly isolated system will never reach equilibrium. But it
does not matter, since true isolation never occurs and external
(``random") disturbances  will always drive the system towards equilibrium
\footnote{One can even invoke a theorem to that effect: the ergodic theorem
for Markov chains. But this is again highly misleading. This theorem says
that probability distributions will converge to an
``equilibrium"`distribution (for suitable chains). This is similar, and
related, to what happens with strongly chaotic systems. But it does not
explain what happens to a single system, unless we are willing to
distinguish between microscopic and macroscopic variables, in which
case the ergodic theorem is not necessary.}.
This is true but irrelevant\footnote{
 Borel \cite{Bor} tried to answer the reversibility objection,
 using the lack of isolation and the
instability of the trajectories. As we saw in Section 3.3, this objection
is not relevant, once one introduces the micro/macro distinction.
And Fred Hoyle
wrote: ``The thermodynamic arrow of time does not come
from the physical system itself\dots it comes from the connection of
the system with the outside world" \cite{Ho}, quoted in \cite{Lan}.
See also  Cohen and Stewart (\cite{CS}, p. 260) and \cite{DH}
for similar ideas.}.

In order
to understand this problem of non-isolation, we have to see how to deal with
idealizations in physics. Boltzmann compares
this with Galilean invariance (see \cite{Bo}, p.170).
Because of non-isolation, Galilean (or Lorentz) invariance can never be
applied strictly speaking (except to the entire universe, which is not very
useful). Yet, there are many phenomena whose explanation
involve Galilean (or Lorentz) invariance. We simply do as if the
invariance was exact and we then
argue that the fact that it is only approximate
does not spoil the argument. One uses a similar reasoning
 in statistical mechanics.
If we can explain what we want to explain (e.g. irreversibility)
by making the assumption that the  system is perfectly isolated,
 then we do not
have to introduce the lack of isolation in our explanations. We have only to
make sure that this lack of isolation does not conflict with our
explanation. And how could it? The lack of isolation should, in general,
speed up the convergence towards equilibrium\footnote{One has to be careful
here. If we shake a mixture of fluids, it should become homogeneous faster.
But of course, there are external influences that prevent the system from
going to equilibrium, as with a refrigerator. Also, the time scale on which
approach to equilibrium takes place may vary enormously, depending
on the physical situation. This is what is
overlooked in \cite{DH}.}. Also, if we want to explain why a
steamboat cannot use the kinetic energy of the water to move, we apply
irreversibility arguments to the system boat$+$water, even though the whole
system is not really isolated.

Another way to see that lack of isolation is true but irrelevant is to
imagine a system being more and more isolated. Is irreversibility going to
disappear at some point? That is, will different fluids not mix themselves,
or will they spontaneously unmix? I cannot think of any example where this
could be argued. And I cannot tell with a straight face to a student that
(part of) our explanation for irreversible phenomena on earth depends on the
{\it existence} of Sirius.

\vs{3mm}

\no {\bf 4.4. Bergson, Popper, Feyerabend (and others).}

\vs{3mm}

Here, I will discuss various confusions that
have been spread by some philosophers.
Bergson was a rather
unscientific thinker, and  many readers may wonder why
 he belongs here. I have myself
been very surprised to see how much sympathy Prigogine and Stengers
seem to have for Bergson (see the references to
Bergson in \cite{PS1,PS2}). But Bergson has been extremely
influential, at least in the
French culture, and, I am afraid, still is\footnote{I remember that
when I first heard, as a teenager, about the special theory of
relativity, it was through Bergson's alleged refutation of that
theory! He thought, probably rightly so, that there was aconflict
between his intuitive views on duration and the absence of absolute
simultaneity in Relativity. So he simply decided that there was a ``time
of consciousness", as absolute as Newtonian time, and that the
Lorentz transformations were merely some kind of coordinates ``attributed"
by one observer to the other. Running into trouble with the twin
paradox, he decided that acceleration is relative, like uniform motion,  and that,
when both
twins meet again, they have the same age! (see \cite{Ber1}).
At least, Bergson had the good sense, after his lengthy
polemic with Einstein, to stop the republication of his
book. But, and this is a remarkable aspect of our ``intellectual"
culture, the {\it very same mistake} is repeated by some of
his admirers, Jankelevitch (\cite{Ya}, Chap. 2),
Merleau-Ponty (\cite{MP}, p. 319) and Deleuze (\cite{Del}, p.79; see also
 his later writings).
Of course, all this is explained  by telling to the physicists that they
should stick
to their ``mathematical expressions and
language" (Merleau-Ponty, \cite{MP}, p. 320), while leaving
the deep problems of the ``time
of consciousness" to philosophers. For a modern attempt to make
some sense
of Bergson's universal time, see the first Appendix of ``Entre
le Temps et l'Eternit\'e" (\cite{PS2}).
}. In particular, he is one source of the widespread confusion
that there is contradiction between life and the
Second Law of thermodynamics. Roughly speaking,
Bergson saw a great opposition between ``matter" and ``life",
and a related one between  intellect and  intuition.
The intellect can understand matter, but intuition is needed
to apprehend life\footnote{See Monod \cite{Mo} and B. Russell
\cite{Rus}  for a
 critique of his philosophy and
\cite{Mor} for the relation between Bergson
and Prigogine. The main problem
with Bergson's lasting
influence
is well expressed by Bertrand Russell: ``One of the bad
effects of an anti-intellectual
philosophy such as that of Bergson, is that it thrives upon the errors
and confusions of the intellect. Hence it is led to prefer bad thinking
to good, to declare every momentary difficulty insoluble, and to regard
every foolish mistake as revealing the bankruptcy of intellect and the
triumph of intuition." (\cite{Rus}, p.831)}. Bergson
was not a precursor of the discovery of
DNA, to put it mildly\footnote{This remark is not as
anachronistic as it may seem. Think of the work of Weismann,
at the turn of the century,
 on the continuity of the germ-plasm.}. The Second Law of
thermodynamics, which he called
the ``most metaphysical of the laws of
physics" (\cite{Ber}, p.264), was very important
for him\footnote{As we shall see in Section 5, it is probably
the least metaphysical of those laws (although I do not like this
terminology), since it is not a purely dynamical law.}. It
reinforced his ``vision of the material world as that
of a falling weight." (\cite{Ber}, p.266), hence, that
``all our analyses show indeed in life an effort to climb
the slope that matter has descended." (\cite{Ber}, p.267)
``The truth is that life is possible wherever energy
goes down the slope of Carnot's law, and where a cause, acting
in the opposite direction, can slow down the descent."(\cite{Ber}, p.278)
It's all  metaphorical, of course, but Bergson's philosophy {\it is}
 entirely a ``metaphorical dialectics devoid of logic,
but not of poetry", as Monod calls it (\cite{Mo}).
In any case, life is perfectly compatible with the Second
Law (see Section 3.3).

Turning to
Popper, we have already seen
 that he had lots of problems with statistical
mechanics. Since Popper
is generally considered positively by
scientists\footnote{See, e.g. the introduction
by Monod to the French edition of
``The Logic of Scientific Discovery" \cite{Pop3} and also
 Prigogine and Stengers, e.g.
(\cite{PS2}, p.173). For philosophical critiques
of Popper, see Putnam \cite{Pu}, and Stove \cite{Sto}.
For a critique of his views on the
arrow of time, see Ghins \cite{Gh}.}, it
is worth looking more closely at his
objections. He took too literally the claims of
Heisenberg, Born and Pauli on irreversibility as ``subjective" (see Section
3.5), which he thought (maybe rightly so)
 were precursors of the subjectivism of the
Copenhagen interpretation of quantum mechanics
(see \cite {Pop4}). Besides, he
was strongly opposed to determinism and he was convinced
that ``the strangely law-like behaviour of the statistical sequences remain,
for the determinist, {\it ultimately irreducible and inexplicable}."
(\cite {Pop5}, p.102). As I discussed in
Section 2.2, there is no problem
in using probabilities, even in a deterministic universe.
He then invented a rather obscure ``propensity" interpretation of
probabilities. He also felt that one should define ``objectively" what
a random sequence is. A sequence (of zeros and ones) will be random
if there are (almost) as many zeroes and ones, as many pairs
$00$, $01$, $10$, $11$, etc\dots
(see e.g. \cite{Pop4}, p.112). He did not seem to realize that
this is like saying that a ``microscopic configuration" (a sequence)
gives to certain ``macroscopic variables" (the average number of occurences
of finite subsequences) the values which are given to them by the
overwhelming majority of sequences. So that the difference with what he
calls the ``subjective" viewpoint is not so great.

Finally, Popper
was very critical of Boltzmann. Although he  admires
Boltzmann's realist philosophy, he calls Boltzmann's interpretation of
time's arrow ``idealist" and claims that it
was a failure.  As we saw, any explanation
of irreversibility ultimately forces us to say that the universe started
in an ``improbable" state. Boltzmann tried to explain  it as follows:
in an eternal and infinite universe globally in equilibrium,
all kinds of fluctuations will occur. What we call our universe
is just the result of one such gigantic fluctuation, on its way back to
equilibrium. But this explanation does not really work.
 Indeed, the most probable assumption, if
a fluctuation theory is to hold, is simply that my brain is a fluctuation
out of equilibrium, just at this moment and
in this small region of space, while
none of the familiar objects of the
universe (stars, planets, other human beings)
 exist and
all my (illusory) perceptions and memories are simply
encoded in the states of
my neurons (a ``scientific" version of solipsism).
However improbable such a fluctuation is, it is still far more
probable than a fluctuation giving rise to the observed universe, of which
my brain is a part.
Hence, according to  the fluctuation theory,
that ``solipsist" fluctuation must
actually have occurred many more times than the big
fluctuation in which we live, and therefore
no explanation is given for
the fact that we happen to live in the latter (see Feynman
\cite{Fe2} and Lebowitz \cite{Le1} for a discussion of that fluctuation
theory).

 Boltzmann's cosmology does not work. So, what?
When Popper wrote (1974), no one took Boltzmann's cosmology
seriously anyway: it had long since
been superseded by cosmologies based
on general relativity.
 Besides, Popper does not raise
the objection I just made. His criticism is, rather,
 that this view would render time's
arrow ``subjective" and make Hiroshima an ``illusion". This is
complete gibberish. Boltzmann gives a complete and straightforward explanation
of irreversible processes in which Hiroshima is as objective as it
 unfortunately is (when it is described at the macroscopic level,
which is what we mean by ``Hiroshima").
Of course, questions remain concerning the initial state of the universe.
In the days of Boltzmann, very little was known about
cosmology. What the failure of
Boltzmann's hypothesis on the origin of the initial state
shows is that cosmology, like the rest of science, cannot be based
on pure thought alone\footnote{There are indications that Boltzmann
did not take his fluctuation theory too seriously. For example, he
wrote ``that the world began from a very unlikely initial state, this much
can be counted amongst the fundamental hypotheses of the whole theory
and we can say that the reason for it is as little known as that for why the
world is as it is and not otherwise."(\cite{Bo}, p.172; compare with
note \protect\ref{note_Feynman}).
In general, Boltzmann is quite opposed to unscientific speculations.
In his criticism of Schopenhauer, he
takes a very Darwinian (and
surprisingly modern) view of mankind. He starts by observing that drinking
fermented fruit juices can be very good for your health: ``if I were an
anti-alcoholic I might not have come back alive from America,
so severe was the dyssentry that I caught as a result of bad water\dots it was
only through alcololic beverages that I was saved."
(\cite{Bo}, p.194) But, with alcohol,
 one can easily
overshoot the
mark. It is the same thing with moral ideas.
 ``We are in the habit of assessing
everything as to its value; according to
 whether it helps or hinders the conditions of life, it
is valuable or valueless. This becomes so habitual that we imagine we must ask
ourselves whether life itself has a value. This is one
of those questions utterly devoid of sense." (\cite{Bo}, p.197)
Finally, for theoretical ideas, he observes that
our thoughts should correspond to experience and
that overshooting the mark should
be kept within proper bounds: ``Even if this ideal will presumably never be
completely realized, we can nevertheless come nearer to it, and this would
ensure cessation of the disquiet and the embarassing feeling that it is a riddle
that we are here, that the world is at all and is as it is, that it is
incomprehensible what is the cause of this regular connection between cause
and effect, and so on. Men would be freed from the spiritual migraine that
is called metaphysics."(\cite{Bo}, p.198).}.

Popper was also too much impressed with
Zermelo's objections to Boltzmann,
based on the Poincar\'e recurrence theorem, and discussed
above (see \cite{Pop1}).
But he
has even stranger criticisms: in \cite{Pop2},
 he argues that Brownian motion (where fluctuations may pull the particle
against gravity) is a serious problem for the Second Law. Maxwell had
already observed that ``The Second Law is constantly being
violated\dots in any sufficiently small group of molecule\dots As the number
\dots is increased \dots the probability of a measurable variation \dots
may be regarded as practically an impossibility." (\cite{Max}, quoted in
\cite{Le2}) Going from bad to worse, Feyerabend invents a ``perpetuum mobile
of the second kind" (i.e. one respecting the first law but not the second)
using {\it a single molecule} \cite{Fe1}.
He adds that he assumes ``frictionless devices" (he had better do so!).
 Those claims are  then repeated in his
popular book ``Against Method" \cite{Fe}, where it is explained that
Brownian motion
refutes the Second Law\footnote{This error is repeated, with
many others, in \cite{Wo}, p.177.}. This is how the general educated public is misled
into believing that there are deep open problems  which are
deliberately ignored by the ``official science"!

Unfortunately, this is not the end of it. Contemporary (or post-modern)
French
``philosophy" is an endless source of confusions on chaos and irreversibility.
Here are just a few examples. The well-known philosopher Michel Serres says, in
an interview with the  sociologist of science Bruno Latour,
entitled paradoxically ``Eclaircissements": ``Le temps ne coule
pas toujours selon une ligne (la premi\`ere intuition se trouve dans un chapitre de mon livre
sur Leibniz, pp.~284--286)
ni selon un plan, mais selon une vari\'et\'e extraordinairement complexe,
 comme s'il montrait des points d'arr\^et,
des ruptures, des puits, des chemin\'ees
d'acc\'el\'eration foudroyante, des d\'echirures,
 des lacunes, le tout ensemenc\'e
al\'eatoirement, au moins dans un d\'esordre visible. Ainsi le d\'eveloppement de
l'histoire ressemble vraiment \`a ce que d\'ecrit la
th\'eorie du chaos \ldots "\footnote{I will leave
these texts in the original language, and provide only a rough
translation, because nonsense is hard to
translate: the book is called ``Clarifications" and the quotation
is: ``Time does not always flow along a line, nor along a plane, but along
an extraordinarily complex manifold, as if it showed stopping points,
ruptures, sinks, chimneys of striking acceleration, rips, lacunas, everything
being randomly sowned, at least in a visible disorder. So, the development of
history really ressembles what is described by chaos theory."} (\cite{Se}).
Another philosopher, Jean-Fran\c cois Lyotard writes:``L'id\'ee que l'on
tire de ces recherches (et de bien d'autres)
est que la pr\'e\'eminence de la fonction continue \`a deriv\'ee comme paradigme de la
connaissance et de la pr\'evision est en train de dispara\^{\i}tre.
En s'int\'eressant aux
ind\'ecidables, aux limites de la pr\'ecision du contr\^ole, aux quanta,
aux conflits \`a l'information non compl\`ete, aux ``{\em fracta}\/'',
aux catastrophes, aux paradoxes pragmatiques, la science postmoderne
fait la th\'eorie de sa propre \'evolution comme discontinue,
catastrophique, non rectifiable, paradoxale.
Elle change le sens du mot savoir, et elle
 dit comment ce changement peut avoir lieu. Elle
produit non pas du connu, mais de l'inconnu. Et elle
sugg\`ere un mod\`ele de l\'egitimation
qui n'est nullement celui de la meilleure
performance, mais celui de la diff\'erence comprise
comme paralogie."\footnote{``The idea derived from those researches
(and from many others) is that the pre-eminence of the continuous
function with a derivative as paradigm of knowledge and forecast
is disappearing. By being interested in undecidables, in limits
of precision of control, in quanta, in conflicts with incomplete information
, in ``fracta", in catastrophes, in pragmatical paradoxes, postmodern
science makes the theory of its own evolution as discontinuous,
catastrophic, not rectifiable, paradoxical. It changes the meaning
of the word knowledge, and it says how this change can occur. It
produces not the known, but the unknown. And it suggests a model
of legitimation which is not at all the one of the best performance,
but rather the one of difference understood as paralogism."}
(\cite{Ly}).
A sociologist, Jean Baudrillard observes
that ``Il faut peut-\^etre consid\'erer l'histoire
elle-m\^eme comme une formation
chaotique o\`u l'acc\'el\'eration met fin \`a la
lin\'earit\'e, et o\`u les turbulences
cr\'e\'ees par l'acc\'el\'eration \'eloignent
d\'efinitivement l'histoire de sa fin, comme
elles \'eloignent les effets de leurs causes. La
destination, m\^eme si c'est le Jugement dernier,
nous ne l'atteindrons pas, nous en sommes
d\'esormais s\'epar\'es par un hyperespace \`a
r\'efraction variable. La r\'etroversion de
l'histoire pourrait fort bien s'interpr\'eter
comme une turbulence de ce genre, due \`a la
pr\'ecipitation des \'ev\'enements qui en inverse le
cours et en ravale la trajectoire."\footnote{``One must,
maybe, consider history
itself as a chaotic formation where acceleration
puts an end to linearity, and where turbulence
created by acceleration separates definitively
history from its end, as it separates effects from their
causes. The destination, even if it is
the Last Judgment, we shall not reach it, we are separated from it
by a hyperspace with variable refraction.
The retroversion of history could very well be
interpreted as such a turbulence, due to the
precipitancy of events which inverts its path and
swallows its trajectory."} (\cite{Bau}).
Finally, Gilles Deleuze and F\'elix Guattari understood  chaos
as follows: `` On
d\'efinit le chaos moins par son d\'esordre que par
 la vitesse infinie avec laquelle se
dissipe toute forme qui s'y \'ebauche. C'est
un vide qui n'est pas un n\'eant, mais un
{\em virtuel}\/, contenant toutes les particules
 possibles et tirant toutes les formes possibles
qui surgissent pour dispara\^{\i}tre aussit\^ot,
sans consistance ni r\'ef\'erence, sans
cons\'equence (Ilya Prigogine et Isabelle Stengers,
{\em Entre le temps et l'\'eternit\'e}\/,
pp.~162--163)."\footnote{``Chaos is defined not so much by its disorder
than by the infinite speed with which
every form being sketched is dissipated.
It is a vacuum which is not a nothingness, but
a {\em virtual}\/, containing all possible particles
and extracting all possible forms, which appear and disappear
immediately, without consistence,
nor reference, nor consequence."} (\cite{DG}) Of course, Prigogine and Stengers are not
responsible for {\it these} confusions (in that reference,
they discuss
the origin of the universe). But this illustrates the difficulties
and the dangers of the popularization of science.
Besides, Guattari wrote a whole book on ``Chaosmose" (\cite{Gu}),
which is full
of  references to
non-existent concepts such as ``nonlinear irreversibility thresholds"
and ``fractal machines"\footnote{I recommand to people interested
in {\it tensors} (applied to psychology, sociology, etc \dots) Guattari's contribution
to \cite{Cer}.}.

\section{Entropies}
\begin{flushleft}
{\footnotesize Holy Entropy! It's boiling!\\
Mr Tompkins (G. Gamow) (\cite{Ga}, p.111).} \end{flushleft}

There is  some kind of mystique about entropy.
According to \cite{Den}, \cite{Tr},  von Neumann
suggested to Shannon to use the word ``entropy" adding that
 ``it will give you a great
edge in debates because nobody really knows what entropy is anyway".
But there is a very simple way to understand the notion of entropy.
Just consider
any set of macroscopic variables (at a given time) and consider
the volume of the
subset of phase space (of the microscopic variables) on which these
macroscopic variables
take a given value. The {\it Boltzmann
entropy} (defined as a function of the values taken by
the macroscopic
variables) equals the logarithm of that volume. Defined this way, it looks
quite arbitrary. We may  define as many entropies as we can find sets of
macroscopic variables. Furthermore, since the micro/macro distinction is not
sharp, we can always take finer grained entropies, until we reach the
microscopic variables (the positions and the momenta of the particles), in
which case the entropy is constant and equals zero (giving a volume equal to
one to a single microstate, which is rather a quantum-mechanical way to
count).

But one should make several remarks:
\begin{enumerate}
\item[1)] These entropies are not necessarily ``subjective". They are as
objective as the corresponding macroscopic variables.
Jaynes, following Wigner, calls these entropies ``anthropomorphic"
(\cite{Ja}, p.85). A better word might be ``contextual", i.e. they depend
on the physical situation and on its level of description.
\item[2)] The ``usual" entropy of Clausius, the one which is most
useful in practice,
corresponds to a particular choice of
macroscopic variables (e.g. energy and
number of particles per unit volume for a
monoatomic gas without external forces). The derivative with respect to the
energy of {\em that} entropy, restricted to equilibrium values, defines the
inverse temperature. One should not
confuse the ``flexible" notion of entropy introduced
above with the more specific one used
in thermodynamics\footnote{This
is again the source of much confusion, just like the ``subjectivity
of irreversibility" discussed in Section 3.5,
see for example Popper \cite{Pop4}, p.111, Denbigh \cite{Den}, S.K. Ma
\cite{Ma1}. In the popular
book, ``The Quark and The Jaguar", we read: ``Entropy and information are
closely related. In fact, entropy can be regarded as a measure of
ignorance."(\cite{Ge}, p.219) And further: `` Indeed, it is mathematically
correct that the entropy of a system described in perfect detail would not
increase; it would remain constant."(\cite{Ge}, p.225).
This is correct, if properly understood,
 but it might be useful to emphasize that one
 does not refer to
 the ``usual" thermodynamic entropy.}.
\item[3)] The Second Law seems now
a bit difficult to state precisely.
``Entropy increases"; yes, but which one?
One can take several attitudes. The most conservative one
is to
restrict oneselve
to the evolution of a given isolated
system between two equilibrium states
and then the increasing
entropy is the one discussed in point (2) above.
The Second law is then
a rather immediate
consequence of the irreversible evolution of
the macroscopic variables: the microscopic motion will
go from small regions of phase space
 to larger ones (in the sense of the partitions discussed in
Section 4.2).  The gas in the box goes from an
equilibrium state in the left half of the box to another equilibrium
state in the whole box. There are many
more microscopic configurations corresponding to a uniform density
than there are configurations corresponding to the gas being entirely in
one half of the box.
But this version of the Second Law
is rather restrictive, since most natural phenomena to which we
apply ``Second Law" arguments are not in equilibrium.
When used properly in non-equilibrium situations, reasonings based
on the Second Law
 give an extremely reliable way to predict how a system will evolve.
We simply assume that a system will never go spontaneously
towards a very small subset of its phase space
(as defined by the macroscopic variables). Hence, if we observe such an
evolution, we expect that some hidden external influence is forcing
the system to do so, and we try to discover it (see also Jaynes \cite{Ja4}
for a nice discussion of apparent
violations of the Second Law)\footnote{See e.g. the
constraints on the plausible mechanisms for the origin of life, due
to the Second Law,
discussed by Elitzur (\cite{El}, Sect.11). There is some similarity
between this use of the Second Law
and the way biologists use the law of natural selection.
The biologists do not believe that complex organs appear ``spontaneously".
Hence, when they occur, they look for an adaptative
explanation (see \cite{Da1} for an introduction to the theory
of evolution).
Both attitudes are of course  similar to elementary probabilistic
reasoning: if we throw a coin a million times and find a significant deviation
from one-half heads one-half tails, we shall conclude that the coin is biased
(rather than assuming that we observe a miracle). }.
\item[4)] In most non-equilibrium situations, most of these entropies are very
hard to compute or even to estimate. However, Boltzmann was able to find an
approximate expression of his entropy (minus his $H$ function),
valid for dilute gases (e.g. for the gas in the box
initially divided in two of
Section 3) and to write down an equation for the evolution
of that approximate entropy. A lot of confusion is
 due to the identification
between the ``general" Boltzmann entropy defined above, and
the approximation to it given by (minus) the $H$-function (as emphasized
by
Lebowitz in \cite{Le1}).
Another frequent confusion about Boltzmann's equation is
to mix two conceptually different
ingredients entering in its derivation\footnote{As for example in:
``This so-called hypothesis of ``molecular chaos" admits the absence of
correlations between the velocities of the molecules in the initial
state of the gas, although, obviously, correlations
exists between the molecules after the collisions. The hypothesis of molecular
chaos amounts to introduce, in a subtle way, the irreversibility that
one tries  demonstrate." (Lestienne \cite{Les}, p.172)
Boltzmann never said that he would demonstrate
irreversibility without assuming
something about initial conditions.
Another, more radical, confusion is
due to Bergmann: ``It is quite obvious that the Boltzmann equation, far
from being a consequence of the laws of classical mechanics, is inconsistent
with them." (in \cite{Go}, p. 191)}: one is
an assumption about {\it initial conditions}
and  the other is to make a particular
 approximation (i.e. one consider
the Boltzmann-Grad limit, see Spohn \cite{Sp}, in which the equation
becomes exact; in the Kac model in Appendix 1,
this limit reduces
simply to letting $n$ go to infinity for fixed $t$). To account for irreversible
behaviour, one has always, as we saw, to assume something
on initial conditions, and the justification of that assumption is
 statistical. But that part does not require, in principle,
 any approximation.
 To write down a concrete
(and reasonably simple) equation, as
Boltzmann did, one uses this approximation. Failure to distinguish these
two steps leads one to believe that there is some deep problem with
irreversibility outside the range of validity of that
 approximation\footnote{To make a {\it vague} analogy, in equilibrium statistical
mechanics, one has the concept of phase transition.
Mean
field theory (or the van der Waals theory, Curie-Weiss
or molecular
field approximation) gives an approximate description of the phase
transition. But the concept of phase transition is much wider than
the range of validity of that approximation.}.
\item[5)] Liouville's theorem\footnote{This theorem says that, if $A$ is a
subset of the phase space $\Omega$,
then $Vol(T^t(A))=Vol(A)$, where $Vol(A)=\int_A d{\bf x}$.
}
is sometimes invoked against such ideas. For instance,
we read in Prigogine and Stengers (\cite{PS2}, p.104):
``All attempts to construct an entropy function, describing the evolution of
a set of trajectories in phase space, came up against Liouville's theorem,
since the evolution of such a set cannot be described by a function that
increases with time"\footnote{See also \cite{Cer}, p.160:
``According to the mechanical
view of the world, the entropy of the universe is
today identical to what it was at the origin of time."
Or as Coveney says (\cite{Co}, p.411): ``As long as the
dynamical evolution is unitary, irreversibility cannot arise.
This is the fundamental problem of non-equilibrium statistical mechanics."}
(see \cite{DH}, p.8 for a similar statement). What is
the solution of that ``paradox"?
Here I consider {\it a single system} evolving in time and associate to it a
certain set of macroscopic variables to which in turn an entropy is
attached. But, since the values of
the macroscopic variables change with time, the
corresponding set of microstates changes too. For the gas in the box,
the initial set of microstates are all those where the particles are
in the left half, while the final set consists of the microstates
giving rise to a uniform density.
In other words, I ``embed" my microscopic state into different
sets of microscopic states as time changes, and the
evolution of that set
 should not be confused with 	a set of {\em trajectories}, whose
volume is indeed forced to remain constant
(by Liouville's theorem)\footnote{Let me use the notations of note
\protect\ref{note_good}.
By Liouville's Theorem, indeed $Vol(T^t(\Omega_0))=Vol(\Omega_0)$.
But $T^t(\Omega_0)$ is a very small subset of $\Omega_t$. Confusing the two
sets leads to the (wrong) idea that
$Vol(T^t(\Omega_0))=Vol(\Omega_t)$. The evolution of $\Omega_t$ does
 not coincide with
a set of trajectories.}.
\item[6)] A related source of confusion comes from the fact that Gibbs'
entropy, $-\int \rho \log\rho d{\bf x}$, which
is sometimes viewed as more ``fundamental" (because
it is expressed
via a distribution function $\rho$ on phase space), is indeed constant in time (by
Liouville's theorem again). But why should one use this Gibbs entropy out of
equilibrium? In equilibrium, it agrees with Boltzmann and Clausius entropies
(up to terms that are negligible when the number of particles is large) and
everything is fine\footnote{Note, that these entropies  agree with (minus)
Boltzmann's $H$ function only when the interparticle forces are negligeable
(as in a very dilute gas). This is rather obvious since the $H$ function is
an approximation to the Boltzmann entropy, see Jaynes (\cite{Ja}, p.81).}.
When we compare
two different equilibrium states all these entropies change, and the
direction of change agrees with the Second Law\footnote{ Amusingly
enough, this conclusion can be reached
using only Liouville's theorem (see Jaynes \cite{Ja} p.83) which is
 blamed as
the source of all the troubles!}. The reason being that the
values taken by
the macroscopic variables are different for different equilibrium
states. Actually, trying to ``force" the Gibbs
entropy to increase by various coarse-graining techniques, gives
then the impression
that irreversibility is only due to
this coarse-graining and is therefore  arbitrary or
subjective (see e.g. Coveney (\cite{Co}, p.412).:
``Irreversibility is admitted into the description by asserting that
we only observe a coarse-grained probability;").
\item[7)] Finally, why should one worry so
much about entropy for non-equilibrium states? A distinction has to be made
between two aspects of irreversibility: one is that macroscopic variables
tend to obey irreversible laws  and the other is that when an isolated
system can go from one equilibrium state to another, the corresponding
thermodynamic entropies are related by an inequality. Both aspects are
connected, of course, and they can be both explained by similar ideas. But
this does not mean that, in order to account for the irreversible behaviour
of macroscopic variables, we have to introduce an entropy function that
evolves monotonically in time. It may be useful or interesting to do so, but
it is not required to account for irreversibility.
 All we really {\it need} is to define suitably
the entropy for equilibrium
states, and that was done a long time ago.
\item[8)] Jaynes rightly says that he does not know what is the
entropy of a cat (\cite{Ja} p.86). The same thing could be said
for a painting,
an eye or a brain. The problem is that there is
no well-defined set of macroscopic variables that is
specified by the expression ``a cat".
\end{enumerate}

\section{Order out of Chaos?
}
\begin{flushleft}
{\footnotesize In my view all salvation for philosophy may be expected\\
to come from Darwin's theory. As long as people believe in a special spirit\\
that can cognize objects without mechanical means, or in a special will\\
that likewise is apt to will that which is beneficial to us, the simplest\\
psychological phenomena defy explanation.\\
L. Boltzmann} (\cite{Bo}, p.193) \end{flushleft}

In this section, I will discuss the ``constructive role"
of irreversible processes\footnote{Note that the word ``chaos"
in the title is  used
in a somewhat ambiguous way: sometimes it has
 the technical meaning of Section 2, sometimes it means ``disordered"
or ``random".}. But I also want to
discuss the impact
of scientific discoveries on the cultural environment. At least since the
Enlightenment and the Encyclopaedia, scientists
have communicated their discoveries to
 society, and, through the popular books and the educational system, have profoundly
influenced the rest of  culture. But one has to be very careful.
In his recent book on Darwin, the philosopher D. Dennett makes a list of
popular misconceptions about the theory of evolution (\cite{De}, p.392).
One of them is
 that one no longer needs the theory of natural selection,
since we have chaos
theory! He does not indicate the precise
source of this strange idea, but this
illustrates
how easily people can be confused by loose talk, analogies and metaphors.

I think that one should clearly reaffirm certain principles: first of all,
no macroscopic system has ever jumped out
of equilibrium spontaneously. Moreover, isolated macroscopic
systems always evolve towards equilibrium. These are
general qualitative statements
that one can make about macroscopic mechanical systems. No
violations of them
have ever been found. Of course, nobody explicitly denies
those principles, but I am nevertheless
afraid that many people
are confused about this point\footnote{Here are some examples; in Cohen and Stewart,
one reads:
``The tendency for systems to segregate into subsystems is just as
common as the tendency for different systems to get mixed together."
(\cite{CS}, p. 259). Or in Meessen: ``by watching
some phenomena, one
is led to say that time has an arrow pointing towards a greater disorder,
but by considering other phenomena, it seems that time has an arrow pointing,
on the contrary, towards a greater order. Then, what does this
arrow mean?
If we can orient it in opposite directions,
it is better not to talk about it any more." (\cite{Me}, p.119).
}.

Of course, it has always been known that
 very complicated and interesting phenomena
 occur out of equilibrium, human beings for example.
But this raises two completely different problems.
Oneis to explain those phenomena on the basis of the microscopic
laws and of suitable assumptions on initial conditions.
Much progress in this direction have been made, but we are far
from understanding everything, and, of course,
to account for the existence of human beings, Darwin's theory
{\it is} needed.

The other question, a much easier one, is to understand why there is
no {\it contradiction} between the general tendency
towards equilibrium
and the  appearance of self-organization, of complex structures
or of living beings.
{\it That} is not difficult
to explain qualitatively, see Section 3.3 and Penrose \cite{Pe2}.

Going back to Popper (again), he wanted to solve the alleged contradiction
between  life and the Second Law (see note
\protect\ref{note_life}) by turning to Prigogine \cite{P1}
and saying that
``{\it open systems in a state far from equilibrium
} show no tendency towards increasing disorder, even though they
produce entropy. But they can export this entropy into their
environment, and can increase rather than decrease their internal
order. They can develop structural properties, and thereby do the
very opposite of turning into an equilibrium state in
which nothing exciting can happen any longer." (\cite{Pop5}, p. 173).
This is correct, provided that part of the environment
{\it is more ordered than the system}, where ``order" is  taken in a
technical sense: the system plus its environment (considered as
approximately isolated) is in a state of low entropy, or is  in a
small subset of its {\it total} phase space and moves towards
a larger subset of that space\footnote{One should always distinguish
this precise but technical sense of order, from our intuitive idea
of order. In particular, when one identifies increase of entropy and
 increase of ``disorder", one should realize that this is correct
only if it is a tautology (i.e. if ``disorder" is defined
through the more precise notion of entropy). Otherwise, it can be
 misleading.
A similar problem occurs with
intuitive words like ``complexity" (or
``information" in the past). If we speak of the
``complexity of the brain", it has of course evolved from a less
``complex" structure, except that those words do not have a precise meaning.
 Note  that precise definitions of complexity,
like algorithmic complexity, do not at all capture the
intuitive meaning of the word, since ``random" sequences
are algorithmically complex, and, whatever ``complexity of the brain"
means, and whatever ``random" means, they do not mean the same thing;
see Gell-mann \cite{Ge}, Chap. 3 for a good discussion of this issue.
} (where the subsets are  elements of a partition like the one
discussed in Section 4.2).
 But it is misleading to suggest that order is created out of nothing,
by rejecting ``entropy" in an unspecified
environment\footnote{Besides, as we saw in Section 5, entropy is not
really a ``substance" to be ``exported". This is a
somewhat strange terminology
for a philosopher like Popper, so critical of ``essentialism".}. It
is not enough
to be an ``open system"; the environment must be in a state of low
entropy.
While it is correct to say that the Second Law ``applies only to
isolated systems", it should not be forgotten that most systems can be
considered, at
least approximately,  as subsystems of isolated ones, and that,
therefore, the Second Law does imply some constraints even for open systems.

Here are some examples which {\it may} create this
confusion\footnote{To avoid misunderstandings, let me stress
that my main criticism here is that one may,
through ambiguous   statements,
 unwillingly mislead the non-specialized reader.}:
In (\cite{Cer}, p.157)
Prigogine wants to give an example
 of one of the ``many phenomena`" that cannot be
understood through the ``general interpretation of
the growth of entropy"  due to Boltzmann.
He considers a system of particles (on a line), which
start in a disordered configuration. Then  ``the strong interactions
between those particles" will push them to form an ordered crystal.
It looks like a ``passage from a disordered situation to an ordered one".
But is this an isolated system? This is not clear if one considers
 the pictures. The final configuration looks like
a perfect crystal. But if there are interactions
between the particles favoring an ordered crystal, the
disordered initial
 configuration
must have been one of high potential energy,
hence the ``ordered" configuration
will have a high kinetic energy, and oscillations will occur. Of course, if
the total initial energy is sufficiently small, the oscillations
will be small and the equilibrium
state will be crystalline. But that is not
incompatible with the ``general interpretation of
the growth of entropy". Equilibrium states maximize
entropy, for a given energy, but may be crystalline (at least
for higher dimensional lattices). This is one example
where maximum entropy is not necessarily the same as
maximal disorder
(in the intuitive sense of the word).
 On the other hand, if
dissipation takes place, the ``passage from a
disordered situation to an ordered one"
is possible, even starting from a configuration of
high potential energy. But this means that some environment
 absorbs the energy of
the system, in the form of heat, hence it increases {\it its} entropy.
And the environment must have been more ``ordered" to start with. Again, this
is in agreement with the ``general interpretation of
the growth of entropy".

To give another example, Prigogine and Stengers
 emphasize in (\cite{PS1}, p.427) that, for the
B\'enard instability\footnote{A fluid is maintained between
two horizontal
plates, the lower one being hotter than the higher one.
If the temperature difference is large enough,
rolls will appear, see e.g. \cite{PS1,PS2}
for a discussion of the B\'enard instability.} to occur {\it one must provide
more heat to the system}.
As noticed by Meessen (\cite{Me}, p.118) ``It is remarkable that the
creation of a structure
is initiated by a source of heat, which is usually a source of disorder".
This quotation shows clearly what is confusing:
 heating suggests an increase of disorder, while
the result is the appearance of a self-organized structure. But what is
needed, of course, is a temperature {\it difference} between the two plates.
So, if one heats up from below, one must have some cooling from above.
The cooling acts like a refrigerator, so it requires some ``ordered" source of
energy. The more one heats, the more efficient must be the cooling.

These are fairly trivial remarks, but which, I believe, have to be made,
at least for the general public, if one wants to avoid giving the impression
that processes violating the Second Law can occur: all the emergence of
complex structures, of whatever one sees, is perfectly compatible with the
universal validity of the ``convergence to equilibrium", provided one
remembers that our universe started (and still is)
in a low entropy state\footnote{Somewhat to my surprise, I found
the following {\it theological} commentary
on the ``constructive role" of irreversible phenomena: ``Each time
that a new order of things appear,
it is marked by the dissipation of a chaotic behaviour, by a broken form
of movement, that is, by a ``fractal", by non-linearity,
by variations or ``fluctuations", by instability and randomness.
In this way  the dynamics of self-organization of matter, which reaches the
great complexity of consciousness,
manifests itself"( Ganoczy, \cite{Gan}, p.79). This is  a bit of an extrapolation,
starting from the B\'enard cells.
The quotation appears in
a section of a chapter on ``God in the language of the physicists",  where
the author refers mostly to Prigogine and Stengers, \cite{PS1,P1}.
}.

Besides, one should be careful with the issue of determinism, at the level
of macroscopic laws, for example when bifurcations occur.
 In many places, Prigogine and Stengers seem to attach
a deep meaning to
 the notion of {\it event}:
``By definition, an
event cannot be deduced from a deterministic law: it implies,
one way or another, that what happened ``could" not have
happened." (\cite{PS2}, p.46)\footnote{And even more misleading:
``In a deterministic world, irreversibility would be
meaningless, since the world
of tomorrow would already be contained in the world
of today, there would be
no need to speak of time's arrow."(\cite{Cer} p.166)
}
Let us consider Buridan's ass. One can describe it as being ``in between"
two packs of food. It could choose
either. But that is a macroscopic description. Maybe one of the eyes of
the ass is tilted in one direction, or some of its neurons are
in a certain state favoring one direction. This is an example
where the macroscopic description does not lead to an autonomous
macroscopic law.
At the macroscopic level, things are indeterminate, and the scheme
of Section 3 does not apply: the microscopic
configurations may fall into different classes, corresponding
to different future evolutions for the macroscopic variables,
and no single class
constitutes an overwhelming majority. Thus,
when we repeat the experiment
(meaning that we control the same {\it macroscopic} variables)
different outcomes will occur, because different experiments will
correspond to microscopic variables that belong to different classes.

The same thing may happen in a
variety of phenomena, e.g. which way a roll
in a B\'enard cell will turn.
But that (true) remark has
 nothing to do with the issue of
determinism, which is meaningful only at the
microscopic level: in a perfectly deterministic universe
(at that level) there will always be lots of situations
where no simple autonomous macroscopic laws can be found, hence
we shall have the illusion of ``indeterminism" if we consider
only the macroscopic level\footnote{It is also a bit too fast to say,
as Prigogine and Stengers do,
that this kind of mechanism allows us to go beyond the ``very old
conflict between reductionists and antireductionists." (\cite{PS1}, p.234,
quoted in \cite{Bou}, p.274)
Any reductionist is perfectly happy to admit that some situations
do not have simple, deterministic, macroscopic descriptions, while
I doubt that antireductionists such as Popper and Bergson would
be satisfied with such a simple admission.}.

One should avoid (once more)
the Mind Projection Fallacy. The macroscopic description
may be all that is accessible to us, hence the future
becomes unpredictable, but, again, it does not mean that Nature
is indeterminate\footnote{Here is another theological commentary:
``Irreversibility means that things happen in time {\it and thanks to time},
that it could be that they did not happen, or did happen otherwise and
that an infinite number of possibilities are always open." ````Inventive"
disorder is part of the definition of things\dots Impredictability which is
not due to our inability to control the nature of things,
but to their nature itself, whose future simply
does  not yet exist, and
could not yet be forecasted, even by ``Maxwell's demon" put on Sirius."
(Gesch\'e, \cite{Ges}, p. 121) The author claims to find his
inspiration on the ``new scientific
understanding of the Cosmos" from, among others,
 ``La nouvelle alliance" (\cite{Ges}, p.120).}.

I will conclude with some remarks on Boltzmann and Darwin, which
may also clarify the relation between
``subjective" evaluations of probabilities and
what we call an ``explanation". As we saw,
Boltzmann had a great admiration for Darwin.
 While preparing this article, I read
in ``La Recherche"  that ``the couple random mutations-selection
has some descriptive value, but not at all an explanatory one" (\cite{Schu}).
That attitude is rather common (outside  biology), but it goes a bit too far.
Actually, there is an analogy between
the kind of explanation given by Darwin and the one given
by Boltzman, and
they are both sometimes similarly misunderstood\footnote{In a critique of several
``almost mystical views of life", which deny ``an evolutionary role
to Darwinian selection", the biologist Elitzur
observes that ``such a misleading discussion of evolution is based
on a complete distortion of thermodynamics" (\cite{El}, p.450).
Besides, I disagree, needless to say, with
the comment of Prigogine and Stengers
 (\cite{PS2}, p.23-24), that there is an ``antithesis"
between Boltzmann and Darwin and that
the theories of Darwin were a success while those
of Boltzmann failed.}
 (of course, Darwin's discovery, although less quantitative
than statistical mechanics, had a much
deeper impact on our culture).
 What does it  mean to explain some fact, like
evolution or irreversibility? As we saw,
 we claim to understand some macroscopically observed behaviour
 when, given some macroscopic
constraint on a system, the overwhelming majority
of the microscopic configurations compatible with those
constraints (and evolving
according to the microscopic laws) drive
the macroscopic variables in agreement with that
observed behaviour.

Turning to Darwin, his problem was to explain the diversity of species
and, more importantly,
 the {\it complexity} of living beings, ``those organs of extreme
perfection and complication", like eyes or brains,
as Darwin called them\footnote{Here, I mean complexity in an intuitive sense.
Of course, it is somewhat related to entropy, because, if we consider
the set of molecules in an eye, say, there are very few ways
 to arrange them so as to produce an eye compared to the number
of arrangement that cannot be used for vision.
But as Jaynes says (see Remark 8 in Section 5), as long as we do
not have a well-defined set of macroscopic variables that define
precisely what an eye is, we cannot give a precise characterization
of this ``complexity" in terms of entropy (and, probably, such an entropy
would not be the right concept anyway).
}.
The fact is that we do not know, and we shall never know
 every microscopic detail about the world, especially about the past
(such as every single mutation, how every animal died etc \dots).
Besides, the initial conditions of the world could be just so
that complex organs are put together in one stroke. To use a common
 image, it would be like ``hurling scrap metal
around at random and happening to assemble an airliner"
(Dawkins,\cite{Da1}, p.8). This does not
violate any known law of physics.
But it would be similar to various ``exceptional" initial conditions
that we encountered before (e.g. the particles going back to the
left half of the box).
And we would not consider an explanation  valid if it appealed
to such ``improbable" initial conditions.
But
to say that such a scenario is ``improbable"
simply means  that, given our (macroscopic)
description of the world, there are very few microscopic configurations
compatible with that description and giving rise to this scenario.
And, indeed, if the world was four thousand years old,  the existence
of those complex organs would amount to a miracle.

To understand the Darwinian explanation, one must take into
account  four
elements, at the level of the macroscopic description:
natural selection (very few animals have offspring),
variation (small differences between parents and offsprings
occur, at least in the long run), heritability and time (the earth is much older than
used to be thought). Then, the claim is that the overwhelming majority of
microscopic events (which mutations occur, which animals
 die whithout children)
compatible with such a macroscopic description
leads to the appearance of those `` organs of extreme
perfection and complication"
\footnote{Not being a biologist, I do not want to enter
into any debate about the origin of life, the speed of evolution, or
how far the Darwinian explanation goes. I only want to underline
the similarity with the type of (probabilistic) explanation
used in statistical physics. As I learned from V. Bauchau \cite{VB}, this analogy
was made already in 1877 by C. S. Peirce :`` Mr.
Darwin proposed to apply the statistical method to biology. The same thing
has been done in a widely different branch of science, the theory of gases.
Though unable to say what the movements of any particular molecule of gas
would be on a certain hypothesis regarding the constitution of this class
of bodies, Clausius and Maxwell were yet able, eight years before the
publicationof Darwin's immortal work, by the application of the doctrine
of probabilities, to predict that in the long run such and such a
proportion of the molecules would, under given circumstances, acquire such
and such velocities; that there would take place, every second, such and
such a relative number of collisions, etc.; and from these propositions
were able to deduce certain properties of gases, especially in regard to
their heat-relations. In like manner, Darwin, while unable to say what the
operation of variation and natural selection in any individual case will
be, demonstrates that in the long run they will, or would, adapt animals to
their circumstances."\cite{Pei}}. Note that we do not need to assume that
mutations are genuinely ``random". They may obey perfectly deterministic laws,
and the randomness may reflect only our ignorance of the details.

A final point which is common to Boltzmann and to Darwin (and his
successors)
is that they  have
 provided ``brilliant confirmations
of the mechanical view of Nature"\footnote{Speaking about DNA
as the solution to
the enigma of ``life", the biologist
Dawkins writes: ``Even those philosophers who had been predisposed
to a mechanistic view of life would not have dared
hope for such a total fulfillment of their wildest
dreams."(\cite{Da}, p.17)
Not surprisingly, Popper said that molecular biology became
``almost an ideology" (\cite{Pop5}, p.172).
As for Bergson, he must be turning over
in his grave.}.
Many people simply cannot swallow mechanical and reductionist
explanations.
They need some vital spirit, some teleological principle
or some other animist view. Their philosophies
 ``thrive upon the errors and
confusions of the intellect".
And this is probably why
 the theories of Boltzmann and of Darwin
have been constantly attacked and
misrepresented. Putting philosophical considerations aside,
I believe that what we understand well, we understand in mechanical
and reductionist terms. There is no such thing as a holist explanation
in science. And thanks to people like Boltzmann and  Darwin
the  ``mechanical view of Nature"
is alive and well, and is here to stay.

 \section{Conclusion: What makes poets happy?}
\begin{flushleft}
{\footnotesize I do not think we should embrace scientific theories because they are
more hopeful, or more exhilarating \dots I feel sensitive on this matter
because, as an evolutionary biologist, I know that people who adopt theories
because they are hopeful finish up embracing Lamarckism, which is false,
although perhaps not obviously so, or Creationism, which explains nothing, and
suggests no questions at all. If non-equilibrium thermodynamics makes
poets happier, so be it. But we must accept or reject it on other grounds."
(\cite{May}, p.257, in a
review of \cite{PS3}).} \end{flushleft}

\vs{3mm}

This paper has been written mainly for scientists. However, many
 references to Prigogine are  found in the literature of
the human sciences  and
 philosophy. But why should anybody in  those fields worry about what
happens in physics or chemistry? In his most recent book \cite{P4}, Prigogine
starts by opposing the objective scientific view of the world with the
subjective view (our feeling of time or of ``free will") which some
philosophers  take as their starting point. His goal is to
reconcile both approaches through his new understanding of physics.

Of course, it would be nice if one could fulfill that goal. But there are
again basic confusions\footnote{See also Maes \cite{Mae}
for a discussion of these problems.}. Take the issue of free will. It is true that if the
fundamental laws of physics are deterministic, and if one rejects dualism, then
free will is, in some sense, an illusion. But it
is not clear that an element of ``intrinsic
randomness" in the fundamental physical laws would make it less an
illusion. The only thing which is clear is that our
inability to predict the future is not very relevant for this discussion.
So that the fact that I am unable to predict which way a B\'enard cell will
rotate is not going to make me  feel ``free".
Ignorance does not explain anything.
And there is no precise sense in which
a ``narrow path" has been found
between ``blind laws" and ``arbitrary events" (\cite{P4}, p.224).

Another confusion concerns the relationship between the
natural and the social sciences.
In our discussion of the macroscopic level versus the microscopic one,
we should locate
the problems
that psychology or the social sciences deal with at a very macroscopic
level. Humans or societies are
 so many scales above molecules that modifications in the
 basic physical  laws is (probably)
almost irrelevant for the understanding
of human actions\footnote{In saying this, I do not want to contradict
a strongly reductionist viewpoint. But higher level laws, even though they
are, in principle,
reducible to the lower level ones, are  not necessarily modified
if the latter change. For example, Navier-Stokes equations were not
much affected by the advent of quantum mechanics.
Besides,
so little is known scientifically
about human actions that establishing a link between what is known there
and molecules is not an urgent problem, to put it mildly. On the other
hand, the word
``almost" is important here. Our knowledge of physics
 does rule out many irrational beliefs
at the human level (see e.g. Weinberg, \cite{We}, p.49
 for further discussion of this point).}. The
main problem of the social sciences is to
exist as sciences, i.e. to discover theories that are well tested and that explain
some non-trivial aspect of human affairs. The only thing that people working in those
fields might learn from the natural sciences is a general scientific attitude, what
one might call the epistemology of the Enlightenment: a critical mind, not to rely
on authorities, to compare theory with experiment, etc\dots. But there is no need to
ape what happens in the exact sciences. So that, even if there was really
a shift of paradigm (whatever that means) from Newtonianism to Prigoginianism
in physics, that would be no reason at all
 for the social scientists to rush towards theories
where randomness is important\footnote{For
an extreme example of this confusion, see \cite{Bec} where quantum theory
is ``applied" to politics.}. And, of course, probabilistic models may be relevant
in the social sciences, even if the fundamental laws are determistic.

The final confusion concerns the ``end of certainties" \cite{P4}
or the ``desillusion with
science" \cite{Bod}. The plain fact
is that we know much more about the world than we did three
centuries ago, or fifty, or twenty years ago. Even the discovery that one cannot
predict the weather (for more than a few weeks) means that our understanding of the laws
governing the weather has improved. The general feeling that there is
 a ``crisis in science" in turn
fuels various anti-scientific attitudes that combine an
 extreme skepticism towards science
with an equally unreasonable openness towards pseudo-sciences
 and superstitions\footnote{See, for example, the discussion
of parapsychology by Stengers in \cite{St}, p.105. She claims that
Rhine, the founder
of parapsychology, has ``devoted all his efforts to invent
increasingly rigorous experimental
protocols, but  meets ``non"-interlocutors, ready to admit any
hypothesis provided it implies that there are no facts". For
a scientific discussion of parapsychology, see e.g. Broch \cite{Bro}. }.
In intellectual circles, this attitude is found in cultural and philosophical relativism or in
some parts of the ``sociology of science"\footnote{I should emphasize that Prigogine
himself does not have an explicit anti-scientific  attitude. But, as  Gross and Levitt
point out in their analysis of  superstitions in academic circles, ``his name keeps coming up
in postmodern discourses with depressing frequency" (\cite{GL}, p. 96). That is
exactly what one would
expect when a famous scientist  tells to
the general educated public, a large part of  which believe in New Age,
in alternative medicines, or in some such nonsense,  that
 one must ``rethink the notion of law of nature". }.
Of course,
science is in a perpetual ``crisis", because it is not a
dogma, and  is subject to revision. But what is not revisable is what I called
the epistemology of the Enlightenment, and I have more than a suspicion that this
epistemology is
 really what is
being attacked by people who insist that there is a deep ``crisis in science".
It is interesting to note (but another article would be needed to develop that point)
that skepticism with respect to science is based on two very different
lines of thought: the first one is based
on traditional philosophical arguments going back to Berkeley, Hume or Kant. While some
of these arguments are clever and interesting, the progress of science is such that these
a priori skeptical arguments leave many people cold. Another, conceptually  different,
line of thought is to try to show that science itself has reached some kind of limit, or
``has to admit"
that one cannot go further. Quantum mechanics, Chaos, the Big Bang or
G\"odel's theorem are usually cited as evidence for those claims. But
this is basically pure confusion
and misunderstanding, as I tried to show in this paper, at least
for one of those examples. When all is said and done,
science and reason is all we have. Outside
of them, there is no hope.

\setcounter{equation}{0}

\vspace{5mm}

{\bf \huge APPENDIX 1. The Kac ring model.}

\vspace{5mm}

Let me analyse a simple model, due to Mark Kac
(\cite{Ka} p.99, see also
Thompson (\cite{Tho} p.23)), which nicely
illustrates Boltzmann's solution to the problem of
irreversibility, and shows how
to avoid various misunderstandings and paradoxes.

I shall describe a slightly modified version of the model
 and state the relevant results,
 referring to \cite{Ka}
for the proofs (the
quotations below
come from \cite{Ka}).

``On a circle we consider $n$ equidistant points"; $m$ of the
 intervals between
the points are marked and form a set called $S$. The complementary set
 (of $n-m$
intervals) will be called
$\bar S$.

``Each of the $n$ points is a site of a ball which can be
either white $(w)$ or
black $(b)$. During an elementary time interval each ball
 moves counterclockwise
to the nearest site with the following proviso".

If the ball crosses an interval in $S$, it changes color
upon completing the move
but if it crosses an interval in
$\bar S$, it performs the move without changing color.

``Suppose that we start with all white balls; the question
is what happens after a
large number of moves".
Below (after eq. 3), we shall also consider other initial conditions.

Let us emphasize the analogy with mechanical laws. The balls are described by
their positions and their (discrete) ``velocity", namely their color. One of the
simplifying features of the model is that the ``velocity" does not affect the
motion. The only reason I call it a
 ``velocity" is that it changes when the ball collides with a fixed
``scatterer", i.e. an interval in $S$. Scattering with fixed objects
tends to be
easier to analyse
than collisions between particles. The ``equations of motion" are given by
the counterclockwise motion, plus the changing of colors (see eqs (5,6) below).
These equations are obviously deterministic and reversible: if after a time $t$, we
change the orientation of the motion from counterclockwise to
clockwise, we return
after $t$ steps to the original state\footnote{There is a small
abuse here, because I seem to change the laws of motion by changing
the orientation. But I can attach another discrete ``velocity" parameter
to the particles, having the same value for all of them, and
indicating the orientation, clockwise or counterclockwise, of
their motion. Then, the motion is truly
reversible, and the operation $I$ of note
\protect\ref{note_involution} simply changes that
velocity parameter.}. Moreover, the motion is strictly periodic:
after $2n$ steps each interval has been crossed twice by each ball, hence they all
come back to their original color. This is analogous to the Poincar\'e cycles,
with the provision that, here, the length of the cycle is the same for all
configurations (there is no reason for this feature to hold in general mechanical
systems). Moreover, it is easy to find special configurations which obviously do
not tend to equilibrium: start with all white balls and
let every other interval
belong to $S$ (with $m=\frac{n}{2}$). Then, after two steps, all balls are black,
after four steps they are all white again, etc... The motion is periodic with
period 4.
Turning to the solution, one can start by analysing the
approach to equilibrium in this model \`a la
Boltzmann:

{\em Analog of the Classical Solution of Boltzmann.} Let $N_w(t)(N_b(t))$ denote
the total number of white (black) balls at time $t$ (i.e., after $t$ moves; $t$
being an integer) and $N_w(S;t)(N_b(S;t))$ the number of white (black) balls
which are going to cross an interval in $S$ at time $t$.

``We have the immediate conservation relations:
\BE
N_w(t+1) &=& N_w(t) - N_w(S;t) + N_b (S;t) \nonumber \\
N_b (t+1) &=& N_b(t) - N_b (S;t) + N_w (S;t) \label{A1}
\EN

Now to follow Boltzmann, we introduce the assumption (``Stosszahlansatz" or
``hypothesis of molecular chaos"\footnote{The word ``chaos" here has nothing
to do with ``chaos theory", and, of course,
Boltzmann's hypothesis is much older than
that theory.})
\BE
N_w(S;t) &=& mn^{-1} N_w (t) \nonumber \\
N_b (S;t) &=& mn^{-1} N_b (t)" \label{A2}
\EN

Of course, if we want to solve (1) in a simple way we have to make some
assumption about $N_w(S;t), N_b (S;t)$. Otherwise, one has to write equations for
$N_w(S;t), N_b (S;t)$ that will involve new variables and lead to a potentially
infinite regress.

The intuitive justification for this assumption is that each ball is
``uncorrelated" with the event ``the interval ahead of the ball   belongs to
$S$", so we write $N_w (S;t)$ as equal to $N_w(t)$, the total number of white
balls, times the density $\frac{n}{m}$ of intervals in $S$. This assumption looks
completely reasonable. However, upon reflection, it may lead to some puzzlement
(just as the hypothesis of ``molecular chaos" does): what does ``uncorrelated"
exactly mean? Why do we introduce a statistical assumption in a mechanical model?
Fortunately here, these questions can be answered precisely and we shall answer
them later by solving the model exactly. But let us return to the Boltzmannian
story.

``One obtains
$$
N_w(t+1) - N_b (t+1) = (1-2mn^{-1})(N_w(t)-N_b(t))
$$
Thus
\BE
n^{-1} [N_w t) - N_b (t)] &=&  (1-2mn^{-1})^t n^{-1}[N_w(0)-N_b(0)]\nonumber\\
&=& (1-2mn^{-1})^t \label{A3}
\EN
and hence if
\be
2m<n
\label{A4}
\en
(as we shall assume in the sequel) we obtain a {\em monotonic} approach to
equipartition of white and black balls." Note that we get a monotonic
approach for {\em all} initial conditions ($N_w(0)-N_b(0)$) of the balls.

The variables $N_w(t), N_b(t)$ play the role of macroscopic variables. We can
associate to them a Boltzmann entropy\footnote{The simplifying
features of the model (the balls do not interact) have
the unpleasant consequence that the ``full"
Boltzmann entropy introduced here and defined in Section 5 actually
coincides with (minus) the Boltzmann $H$-function. But, in general, the
latter should only be an approximation to the former.} $S_b = ln \left( \begin{array}{c} n \\
N_w(t) \end{array} \right)$, i.e. the logarithm of the number of (microscopic)
configurations whose number of white balls is $N_w(t)$. Since
$$
\left( \begin{array}{c} n \\
N_w(t) \end{array} \right) =\frac{n!}{N_w(t)! (n-N_w(t))!}
$$
reaches its maximum value for $N_w = \frac{n}{2} = N_b$,
we see that (3) predicts
a monotone increase of $S$ with time. We can also introduce a partition
of the ``phase space" according to the different values of $N_w$, $N_b$.
And what the above formula shows is that different elements of the partition
have very different number of elements, the vast majority
corresponding to  ``equilibrium", i.e. to those
 near $N_w = \frac{n}{2} = N_b$.

We can see here in what sense Boltzmann's solution is an approximation. The
assumption (2) cannot hold for all times and for all
configurations, because it
would contradict the reversibility and the periodicity of the motion. However,
we can also see why the fact that it is an approximation does not invalidate
Boltzmann's ideas about irreversibility.

Let us reexamine the model at the microscopic level, first
mechanically and then
statistically. For each $i=1,\cdots,n,$ we introduce the variable
\bea
\epsilon_i = \left\{ \begin{array}{c} +1 \; \mbox{if the interval in front of} \; i
\in
\bar S
\\ -1 \; \mbox{if the interval in front of} \; i \in S \end{array} \right.
\eea
and we let
\bea
\eta_i (t)  = \left\{ \begin{array}{c} +1 \; \mbox{if the ball at site} \; i \;
\mbox{at time} \;  t \; \mbox{is white}\\
-1 \; \mbox{if the ball at site} \; i \;
\mbox{at time} \;  t \; \mbox{is black} \end{array} \right.
\eea
Then, we get the ``equations of motion"
\be
\eta_i (t) = \eta_{i-1} (t-1) \epsilon_{i-1}
\label{A5}
\en
whose solution is
\be
\eta_i (t) = \eta_{i-t} (0) \epsilon_{i-1} \epsilon_{i-2} \cdots \epsilon_{i-t}
\label{A6}
\en
(where the subtractions are done modulo $n$). So we have an explicit solution of
the equations of motion at the microscopic level.

We can express the macroscopic variables in terms of that solution:
\be
N_w (t) - N_b(t) = \sum_{i=1}^n \eta_i (t) = \sum^n_{i=1} \eta_{i-t} (0)
\epsilon_{i-1} \epsilon_{i-2} \cdots \epsilon_{i-t}
\label{A7}
\en
and we want to compute $n^{-1} (N_w (t) - N_b (t))$ for large $n$, for various
choices of initial conditions $(\{ \eta_i (0)\})$ and various sets $S$
(determining the $\epsilon_i$'s). It is  here that ``statistical" assumptions
enter. Namely, we fix an arbitrary initial condition $(\{ \eta_i (0)\})$
 and consider all possible
sets $S$ with $m=\mu n$ fixed (one can of course think of the choice of $S$ as
being part of the choice of initial conditions). Then, for each set $S$, one
computes the ``curve" $n^{-1} (N_w (t) - N_b (t))$ as a function of time. The
result of the computation, done in \cite{Ka}, is that, for any given $t$ and for
$n$ large, the overwhelming majority of these curves will
approach $(1-2\frac{m}{n})^t=(1-2\mu)^t$, i.e. what is predicted by (3).
(to fix ideas, Kac suggests to think of $n$ as being of the order $10^{23}$ and
$t$ of order
$10^6$). The fraction of all curves that will deviate significantly from
$(1-2\mu)^t$, for fixed  $t$, goes to zero as $n^{-\frac{1}{2}}$, when $n\to
\infty$.

Of course when I say ``compute" I should rather say that one makes an estimate of
the fraction of ``exceptional" curves deviating from $(1-2\mu)^t$ at a fixed $t$.
This estimate is similar to the law of large number and (7) is indeed of the form
of a sum of (almost independent) variables.

\vspace{3mm}
{\bf Remarks}
\begin{enumerate}
\item[1.]
The Poincar\'e recurrence and the reversibility ``paradoxes" are easily solved:
each curve studied is periodic of period $2n$. So that, if we did not fix $t$ and
let $n\to \infty$, we would not observe ``irreversible" behaviour. But this limit
is physically correct. The recurrence time $(n)$ is enormous compared to any
physically accessible time. As for the reversibility objection,
 let us consider as initial
condition a reversed configuration after time $t$. Then we know that, for that
configuration and {\it that set} $S$, $n^{-1} (N_w(t)-N_b(t))$ will not be close to
$(1-2\mu)^t$ at time $t$ (since it will be back to its initial value $1$).
 But all we are saying is that, for the vast majority of
$S$'s this limiting behaviour will be seen. For the reversed configuration, the
original set $S$ happens to be exceptional. The same remark holds for the
configuration with period 4 mentioned in the beginning.

Note also that, if we consider the set of configurations for which $n^{-1}
(N_w(t)-N_b(t))$ is close to $(1-2\mu)^t$ for {\em all times},
then this set is
empty, because of the periodicity.
\item[2]
We could consider other macroscopic variables, such as the number of white and black
balls in each half of the circle $(1 \leq i \leq \frac{n}{2}$ and $\frac{n}{2} +
1 \leq i \leq n)$, and define the corresponding entropies. We could go on, with
each  quarter of the circle etc..., until we reach a microscopic configuration
(number of white or black ball at each site) in which case the entropy is
trivially equal to zero (and therefore constant).
\item[3] This model, although perfectly ``irreversible", is not ergodic! Indeed
since it is periodic, no trajectory can ``visit" more than 2$n$ microscopic
configurations. But the ``phase space" contains $2^n$ configurations (two
possibilities -black or white- at each site). So, only a very small fraction of
the phase space is visited by a trajectory. This nicely illustrates the fact that
ergodicity is not necessary for irreversibility. What is used
here is
 only the fact that the vast majority of configurations
give to the macroscopic
variables a value close to their equilibrium one.

\end{enumerate}

\vspace{3mm}
{\bf Conclusion}
\vspace{3mm}

I do not want to overemphasize the interest of this model. It
has many simplifying features (for example, there is no conservation
of momentum; the scatterers here are ``fixed", as in the Lorentz gas).
However, it has $all$ the properties that have been invoked to show that
mechanical systems cannot behave irreversibly, and therefore it is a
perfect counterexample that allows us to refute all those arguments (and to
understand exactly what is wrong with them): it is isolated (the balls plus
the scatterers), deterministic, reversible, has Poincar\'e cycles and is not
ergodic.

This result, obtained in the Kac model, is exactly what one would
like to show for general mechanical systems,
in order to establish irreversibility. It is obvious why this is very hard. In
general, one does not have an explicit solution (for an $n$-body system !) such as
(5,6), in terms of which the macroscopic variables can be expressed, see (7).
It
is also clear in this example what is exactly the status
of our ``ignorance". If
we prepare the system many times and if the only variables that we can control
are $n$ and $m$, then we indeed expect to see the irreversible behaviour obtained
above, simply because this is what happens {\it deterministically} for the vast
majority of microscopic initial conditions corresponding to the
macroscopic variables that we are able to
control. We may, if we wish, say that we ``ignore" the initial conditions, but
there is nothing ``subjective" here.
Finally, I shall refer to Kac \cite{Ka}, for
a more detailed discussion, in this model, of
the status of various approximations
used in statistical mechanics (e.g. the Master equation).
\setcounter{equation}{0}

\vspace{5mm}

{\bf \huge APPENDIX 2. On Spectral Representations.}

\vspace{5mm}

I will briefly discuss the mathematical basis of the claim that
``trajectories are eliminated from the probabilistic description." The relevant
mathematics are nicely summarized in the Appendix of \cite{P2} and I shall
therefore refer to that Appendix. Let $T$ be an invertible transformation on a
space $X$ and let $\mu$ be a measure invariant under $T$. In \cite{P2}, $X$ is
the unit square, $T$ the baker's map and $\mu$ is the Lebesgue measure. We can
associate to $T$ a unitary operator $U$ in $L^2$ ($X,\mu)$ (or an isometry
in any $L^p
(X,\mu)$):
\be
Uf (x) = f(T^{-1} x)
\en
and $U^\dagger = U^{-1}$ is
defined\footnote{I follow here the conventions of \cite{P2} for the definition of
$U,U^\dagger$. In \cite{P2} non-invertible transformations such as the Bernoulli
map are also considered. $U$ describes here the evolution of probability
distributions and is  called the Perron-Frobenius operator.}
 by
\be
U^\dagger f(x) = f(Tx)
\en

Since the operators $U$ and $U^\dagger$ are {\em entirely defined} in terms of
$T$, it seems bizarre, to put it middly, to claim that {\em any} property of $U$,
for example its spectral properties, are ``irreducible" to trajectories (i.e. to
the action of $T$). ``Irreducible" is a semi-philosophical notion so that one has
some freedom in the way one uses this word, but I do not think that the meaning
of the word in this context is close to what the general educated public has in
mind\footnote{Compare with statements such
as: ``The mind is irreducible to the body" or ``The behaviour
of a crowd is irreducible to the psychology of individuals".}.

Anyway, putting this issue aside, one should note that the operator $U$ can be
defined more generally. For example, $U$ acts on distributions. In particular, writing
$\delta_{x_0} (x) = \delta (x-x_0)$, we have, for a volume preserving map, like
the baker's map,
\be
U \delta_{x_0} (x) = \delta_{Tx_0} (x)
\en
and the action of $U$ on such delta functions just reflects the evolution of
trajectories.

Another observation is that it has been known for some time that properties of
the dynamics (of the map $T$), are reflected in spectral properties of $U$: for
example, $T$ is ergodic if and only if 1 is a non-degenerate eigenvalue of $U$.

The new feature discussed in \cite{P2} is that for chaotic systems such as the
bakers's map one can write down a spectral representation of the form:
\be
U = \Sigma | F_n (x,y) \rangle 2^{-n} \langle \widetilde{F_n} (x,y)|
\en
where $F_n$ (resp. $ \widetilde{F_n}$) is a product of a polynomial in $x$ (resp. in
$y$) times a distribution in $y$ (resp. in $x$).

This fact is very
 interesting mathematically but it is difficult to see why it implies
radical consequences on ``the laws of nature". The argument given in \cite{P2} is
that the representation (4) cannot be applied to delta functions
 in $x$ and $y$, which
would represent a point, or a trajectory, because  $\widetilde{F_n}$ involves a
distribution in $x$ and one cannot multiply distributions.

Let us see the force of this argument. Here is an analogy. Consider the operator $
 \frac{d}{dx}$. In $L^2 ({\bf R}, dx)$ its spectrum is the imaginary axis and one
can write
\be
\frac{d}{dx} f(x) = \frac{1}{\sqrt{2\pi}} \int i k e^{ikx} \hat f (k) dk
\en
where $\hat f(k)$ is the Fourier transform of $f$. Now obviously for this formula to
hold, it is necessary that $\hat f (k)$ exists. But it is easy to find functions
that are differentiable but whose Fourier transform does not exist. It would be
strange to say that, for those functions, derivatives are eliminated from the
irreducible representation (5). And, of course, as we have seen in (3) the operator
$U$ can perfectly be defined on delta functions. Simply the formula (4) does not
apply in that case.

It is also claimed that formula (4) includes the ``approach to
equilibrium". But, as we discussed in Section 3, the notion of equilibrium does not
make sense for a system with a single degree of freedom, so that this would be one
more argument, if any more were needed, if favour of a dynamics expressed
{\it fundamentally} in terms of trajectories.

All this illustrates the remark made by J.T. Schwartz in his severe critique of the
``pernicious influence of mathematics on science": ``The intellectual attractiveness
of a mathematical argument, as well as the considerable mental labor involved in
following it, makes mathematics a powerful tool of intellectual prestidigitation - a
glittering deception in which some are entrapped, and some, alas,
 entrappers."
\cite{Sch}.

\vs{3mm}

\no{\Large\bf Acknowledgments}

\vs{3mm}

I have discussed many of the issues
raised in this paper with  colleagues and students,
and particularly
with S. Goldstein, A. Kupiainen, J. L. Lebowitz, C. Maes,
J. Pestieau, O. Penrose,
and H. Spohn. I thank
I. Antoniou, B. Misra and I. Prigogine for discussions
on a preliminary draft of this paper.
I have also benefited from discussions with V. Baladi,
V. Bauchau, S. Focant, M. Ghins, L. Haine,
N. Hirtt, D. Lambert, R. Lefevere, I. Letawe, E. Lieb, J.-C. Limpach,
T. Pardoen, P. Radelet  P. Ruelle and E. Speer.

\vs{3mm}

\end{document}